\documentclass[structabstract]{aa}
\usepackage{natbib}
\usepackage[T1]{fontenc}
\usepackage[version=3]{mhchem}
\usepackage{times}
\usepackage{amsmath}
\usepackage{graphicx}

\newcommand{\sub}[1]{_{\rm #1}}
\newcommand{\CII}{[C\,{\sc ii}]}   
  
\newcommand{\thirteenCII}{[$^{13}$C\,{\sc ii}]}
\newcommand{\thirteenCI}{[$^{13}$C\,{\sc i}]}

\newcommand{\HII}{H\,{\sc ii}}

\newcommand{\emm}[1]{\ensuremath{#1}}   
\newcommand{\emr}[1]{\emm{\mathrm{#1}}} 
\newcommand{\unit}[1]{\emm{\, \emr{#1}}}

\newcommand{\kms}{\unit{km\,s^{-1}}}

\renewcommand{\deg}{\emm{^\circ}}
\newcommand{\chr}{^{\rm h}}
\newcommand{\cmin}{^{\rm m}}
\newcommand{\csec}{^{\rm s}}

\newcommand{\FR}{\mbox{\it FR}}
\newcommand{\IR}{\mbox{\it IR}}
\newcommand{\changed}{}
\newcommand{\changedi}{}
\newcommand{\changedii}{}
\makeatletter
\newcounter{reaction}
\renewcommand\thereaction{C\,\arabic{reaction}}
\newcommand\reactiontag{\refstepcounter{reaction}\tag{\thereaction}}
\newcommand\reaction@[2][]{\begin{equation}\ce{#2}%
\ifx\@empty#1\@empty\else\label{#1}\fi%
\reactiontag\end{equation}}
\newcommand\reaction@nonumber[1]{\begin{equation*}\ce{#1}%
\end{equation*}}
\newcommand\reaction{\@ifstar{\reaction@nonumber}{\reaction@}}
\makeatother

\begin{document}
\def\aj{AJ}%
\def\actaa{Acta Astron.}%
\def\araa{ARA\&A}%
\def\apj{ApJ}%
\def\apjl{ApJ}%
\def\apjs{ApJS}%
\def\ao{Appl.~Opt.}%
\def\apss{Ap\&SS}%
\def\aap{A\&A}%
\def\aapr{A\&A~Rev.}%
\def\aaps{A\&AS}%
\def\azh{AZh}%
\def\baas{BAAS}%
\def\bac{Bull. astr. Inst. Czechosl.}%
\def\caa{Chinese Astron. Astrophys.}%
\def\cjaa{Chinese J. Astron. Astrophys.}%
\def\icarus{Icarus}%
\def\jcap{J. Cosmology Astropart. Phys.}%
\def\jrasc{JRASC}%
\def\mnras{MNRAS}%
\def\memras{MmRAS}%
\def\na{New A}%
\def\nar{New A Rev.}%
\def\pasa{PASA}%
\def\pra{Phys.~Rev.~A}%
\def\prb{Phys.~Rev.~B}%
\def\prc{Phys.~Rev.~C}%
\def\prd{Phys.~Rev.~D}%
\def\pre{Phys.~Rev.~E}%
\def\prl{Phys.~Rev.~Lett.}%
\def\pasp{PASP}%
\def\pasj{PASJ}%
\def\qjras{QJRAS}%
\def\rmxaa{Rev. Mexicana Astron. Astrofis.}%
\def\skytel{S\&T}%
\def\solphys{Sol.~Phys.}%
\def\sovast{Soviet~Ast.}%
\def\ssr{Space~Sci.~Rev.}%
\def\zap{ZAp}%
\def\nat{Nature}%
\def\iaucirc{IAU~Circ.}%
\def\aplett{Astrophys.~Lett.}%
\def\apspr{Astrophys.~Space~Phys.~Res.}%
\def\bain{Bull.~Astron.~Inst.~Netherlands}%
\def\fcp{Fund.~Cosmic~Phys.}%
\def\gca{Geochim.~Cosmochim.~Acta}%
\def\grl{Geophys.~Res.~Lett.}%
\def\jcp{J.~Chem.~Phys.}%
\def\jgr{J.~Geophys.~Res.}%
\def\jqsrt{J.~Quant.~Spec.~Radiat.~Transf.}%
\def\memsai{Mem.~Soc.~Astron.~Italiana}%
\def\nphysa{Nucl.~Phys.~A}%
\def\physrep{Phys.~Rep.}%
\def\physscr{Phys.~Scr}%
\def\planss{Planet.~Space~Sci.}%
\def\procspie{Proc.~SPIE}%
\let\astap=\aap
\let\apjlett=\apjl
\let\apjsupp=\apjs
\let\applopt=\ao

\title{Herschel/HIFI observations of \CII{} and \thirteenCII{} in PDRs\thanks{Herschel is an ESA space observatory with science instruments provided by European-led Principal Investigator consortia and with important participation from NASA.}}
\titlerunning{\CII{} and \thirteenCII{} observations in PDRs}
\author{V.~Ossenkopf\inst{1}, M.~R\"ollig\inst{1}, D.A.~Neufeld\inst{2}, P.~Pilleri\inst{3,4}, D.C.~Lis\inst{5}, A.~Fuente\inst{3},
F.F.S.~van der Tak\inst{6,7}, E.~Bergin\inst{8}} 
\authorrunning{Ossenkopf et al. }
\institute{I. Physikalisches Institut, Universit\"at zu K\"oln, Z\"ulpicher Str. 77, D-50937 K\"oln,
Germany\\ \email{ossk@ph1.uni-koeln.de}
\and
Department of Physics and Astronomy, Johns Hopkins University, 3400 North Charles Street, Baltimore, MD 21218, USA
\and
Observatorio Astron\'omico Nacional, Apdo. 112, E-28803 Alcal\'a de Henares (Madrid), Spain	
\and
Centro de Astrobiolog\'{\i}a (INTA-CSIC),
             Ctra. M-108, km.~4, E-28850 Torrej\'on de Ardoz, Spain
\and
California Institute of Technology, Cahill Center for Astronomy and Astrophysics 301-17, Pasadena, CA 91125 USA
\and
SRON Netherlands Institute for Space Research, Landleven 12, 9747 AD Groningen, The Netherlands
\and
Kapteyn Astronomical Institute, University of Groningen, PO box 800, 9700 AV Groningen, The Netherlands
\and
University of Michigan, Ann Arbor, MI 48197 USA
}

\abstract
{Chemical fractionation in the interstellar medium can create
isotopologue abundance ratios that differ by many orders of magnitude
from the normal isotopic abundance ratios. {\changedi By understanding 
how these differences are reflected in astronomical observations under 
various conditions, we try to obtain a new tool \rm}
to sensitively probe these conditions.}
{Recently, we introduced detailed isotopic chemistry into the 
KOSMA-$\tau$ model for photon-dominated regions (PDRs) to give 
theoretical predictions for the abundance of the carbon isotopologues
as a function of PDR parameters. {\changed Combined with radiative 
transfer computations for specific geometries, we estimated the
possible intensity ratio of the \CII{}/\thirteenCII{} lines.
Here, we compare these predictions} with new Herschel observations.}
{We performed Herschel/HIFI observations of the \CII{} $158~\mu{}$m line in
a number of PDRs. {\changed In all sources we observed at least two
hyperfine components} of the \thirteenCII{} transition
allowing to determine the \CII{}/\thirteenCII{} intensity ratio,
after some revision of the {\changed intrinsic hyperfine ratios. 
Comparing the intensity ratios with the results from the updated 
KOSMA-$\tau$ model, we identify cases dominated by chemical 
fractionation and cases dominated by the optical depth of the
main isotopic line}.}
{An observable enhancement of the \CII{}/\thirteenCII{} intensity ratio 
due to chemical fractionation depends mostly on {\changedi geometry and velocity
structure, \rm} and less
on the gas density and radiation field. It is expected to be at maximum for 
{\changed PDR layers that are somewhat shielded from UV radiation, but
not hidden by surface layers of optically thick \CII{}.
In our observations the \CII{}/\thirteenCII{} ratio for the integrated
line intensity was always dominated by the
optical depth of the main isotopic line. However, an enhanced intensity ratio
is \rm} found for particular velocity components {\changed in a few sources:
the red-shifted material in the ultracompact \HII{} region 
Mon~R2, the wings of the turbulent profile in the Orion Bar, and 
possibly a blue wing in NGC~7023 \rm}. The mapping of the
\thirteenCII{} lines in the Orion Bar allows to derive a C$^+$ column
density map confirming the temperature stratification of the C$^+$ layer,
in agreement with the chemical stratification of the Bar. }
{
Carbon fractionation can be significant, even in relatively warm PDRs,
but a resulting enhanced \CII{}/\thirteenCII{} intensity ratio is
only observable for {\changedi special configurations}. In most
cases, a reduced {\changedii \CII{}/\thirteenCII{} intensity ratio \rm}
can be used instead to derive
the \CII{} optical depth to achieve reliable column density estimates
{\changed that can be compared with PDR models}. The C$^+$
column densities for all sources show that at the position of the \CII{}
peak emission, a dominant fraction of the gas-phase carbon is in the form
of C$^+$.
}
\keywords{ISM: abundances -- ISM: structure -- ISM: clouds}

\maketitle

\section{Introduction}

The isotopic abundance ratios of elements such as C, N and O in the 
interstellar medium reflect the star formation history of the environment,
and observations of molecular isotopic line ratios at (sub)mm wavelengths
are a common way to measure this effect \citep[e.g.][]{Wilson1999}. 

Isotopic fractionation of carbon in the interstellar medium, i.e
the $^{12}$C/$^{13}$C ratio in various species, is driven mainly
through the fractionation reaction
\reaction{^{13}C+ + CO <=> C+ + ^{13}CO + {\rm 34.8} \,{\rm K}\label{13eq1}}
providing a significant energy excess for binding the $^{13}$C to the oxygen
\citep{langer84}.
Thus chemical fractionation of carbon is important even for relatively warm
gas, as long as there is a significant fraction of ionized carbon 
available\footnote{All other fractionation reactions for $^{13}$C have
lower excess energies, therefore being relevant only in cold gas.}.
The reaction~(\ref{13eq1}) favors formation of $^{13}$CO and depletion of $^{13}$C$^+$
at low temperatures, shifting the fractional abundances {\changed of the
two species} relative to the
normal solar isotopic ratio, but competes with isotope-selective
dissociation governed by a different shielding through dust, H$_2$ and 
other CO {\changed isotopologues} \citep{Visser2009}.

A systematic discussion of the observational results on carbon
fractionation was provided by \citet{Keene1998} for dense molecular clouds
and \citet{liszt07} with emphasis on more diffuse clouds.
{\changedii Through observations of} \thirteenCI{} and $^{13}$C$^{18}$O in the Orion Bar
\citet{Keene1998} found little evidence for chemical fractionation, 
but noticed a contradiction between the chemical models
predicting a slight relative enhancement of $^{13}$C, but a reduction
of $^{13}$C$^{18}$O while the observations showed the opposite
behavior. {\changed In contrast to the CO results, \citet{Sakai2010} found
clear evidence for fractionation in C$_2$H under the dark cloud
conditions of TMC~1. In line with previous results \citep{Sonnentrucker2007,
sheffer2007,Burgh2007} \citet{liszt07} summarized the clear indications
for fractionation of CO in translucent and diffuse clouds. They
confirmed that the fractionation
reaction~(\ref{13eq1}) dominates the carbon chemistry for densities
$n\sub{H_2} \ge 100$~cm$^{-3}$ and provided new reaction rates for low
temperatures, but} also showed the importance of the CO destruction
in reactions with He$^+$ for lower densities. 

In paper I \citep{paperI}, we present results from the update of
the KOSMA-$\tau$ model for photon-dominated regions (PDRs) to the full isotopic chemistry, providing
isotopologue ratios for various species. Comparing the fractionation
ratio of the $^{12}$C$^+$/$^{13}$C$^+$ abundances with the elemental isotopic 
ratio, we note that
\begin{itemize}
\item{}the fractionation ratio is always higher than or equal to the elemental ratio, i.e. \ce{^{13}C+} is always underabundant with respect to \ce{C+}.
\item{}the fractionation ratio equals the elemental ratio at low $A_V$, but increases significantly towards larger $A_V$.
\end{itemize}
These are  direct consequences of reaction~(\ref{13eq1}). No other important
reaction or isotope-selective photodestruction acts in the opposite direction.
At low $A_V$, temperatures are usually high enough so that the excess energy
plays no role, but once the temperature drops below 50~K the back reaction
becomes less probable and the fractionation ratio increases rapidly. It stays at high values
as long as enough \ce{^{13}C+} ions are available to feed the reaction. 
Deep into the cloud interiors, these ions are formed through the dissociative reaction
\ce{He+ + ^{13}CO}.  The fractionation ratio is then controlled by cosmic ray ionization 
providing a roughly constant $^{12}$C$^+$/$^{13}$C$^+$ enhancement. 

{\changedi However, two competing processes change the \CII{}/\thirteenCII{}
ratio in PDRs: chemical fractionation raises the \CII{}/\thirteenCII{} intensity
ratio relative to the elemental abundance ratio, while the optical depth of the
main isotopic line lowers its intensity relative to the \thirteenCII{} 
transitions.}

{\changed \thirteenCII{} was first resolved by early heterodyne observations at the 
Kuiper Airborne Observatory \citep[KAO, e.g.][]{Boreiko1988,BoreikoBetz1996}
showing no increase of the corresponding \CII{}/\thirteenCII{}
intensity ratio, but a reduction relative to the isotopic elemental ratio
{\changedi corresponding to a significant optical depth of the main isotopic 
line reducing the intensity ratio. With the low spatial and velocity resolution
of these observations, it was, however, impossible to determine wether
this average reduction represents all density or velocity components in the
beam or whether individual components are dominated by either of the two 
competing processes. Only from additional observations with high velocity 
resolution preferentially covering varying conditions} it is therefore possible
to separate the impact of the two processes in PDRs.

The} HIFI instrument \citep{deGraauw} on-board the Herschel
satellite \citep{Pilbratt} allowed for the first time to systematically
observe \thirteenCII{} with a good sensitivity and high spatial
resolution in a number of objects. {\changed This allows to study the $^{12}$C/$^{13}$C
ratio in different environments by comparing the observations with the 
theoretical predictions.
We detected \thirteenCII{} in four bright PDRs, allowing to look for the 
signature of chemical carbon fractionation in the warm PDR gas or to
determine the optical depth and column density of C$^+$.}

In Sect.\,2 we elaborate on the detailed theoretical predictions 
for {\changed $^{12}$C$^+$/$^{13}$C$^+$ and the resulting intensity
ratio \CII{}/\thirteenCII{} under varying physical conditions}
in the frame of the KOSMA-$\tau$ PDR model. 
In Sect.\,3 we present the
HIFI observations of \CII{} and \thirteenCII{}. Section 4 discusses the
observational results for the Orion Bar, Mon~R2, NGC~3603, Carina, and NGC~7023
and Sect.\,5 concludes with a summary of the matches and discrepancies.

\section{Model predictions for the \CII{}/\thirteenCII{} ratio}
\label{sect_model}

{\changed The development of theoretical models for PDRs is an
active field of research because of the complex interplay of
geometrical constraints, properties of the UV field, and chemical
and physical processes in the gas \citep{comparison07}. Because 
of this complexity, all models with an elaborated description of
the PDR chemistry are currently restricted to one-dimensional geometries,
{\changedii therefore describing PDRs} either as plane-parallel slabs or as
(ensemble of) spheres. However, real sources like the prototypical
PDR of the Orion Bar, show more complex geometries, containing
elements of both descriptions, e.g. a large-scale plane-parallel
configuration superimposed on a clumpy small-scale structure, better
described by an ensemble of spherical cores \citep[see e.g.][]{Hogerheijde1995}.}

{\changed We consider both limiting cases in the frame of the
KOSMA-$\tau$ PDR model, a code designed to simulate spherical clumps,
where we can approximate a plane-parallel PDR by considering very massive 
clumps ($M \ge 100 M_{\sun}$) that form an almost plane-parallel
structure in their outer layers. This approach allows us to use
the new implementation of the fractionation network in the 
KOSMA-$\tau$ PDR model code and to easily compare the results
with other PDR models \citep[see][]{comparison07} that lack
isotope networks so far.}

\subsection{Model setup}

The KOSMA-$\tau$ PDR model assumes a spherical cloud geometry
exposed to isotropic UV illumination, measured in units of the 
interstellar radiation field, $\chi_0$ \citep{draine78}. Model
parameters are the hydrogen density at the surface, $n\sub{surf}$, UV field, $\chi$, and 
cloud mass, $M$.
The density structure assumes a power law increase, 
$n_H(r)= n\sub{surf} (r/R_0)^{-1.5}$, from the surface to 1/5 of the outer 
radius, $R_0$,  and a constant density further in.
For the results presented here, we keep the metallicity at solar
values and assume a standard elemental isotopic ratio of
$^{12}$C/$^{13}$C of 67 \citep{wakelam08}. The actual
elemental isotopic ratio is subject to systematic variations with galactocentric radius
\citep{Wilson1999} but all the sources discussed in 
Sect.~\ref{sect_observations} have approximately solar 
galactocentric radii. However, even within the solar neighborhood
there exist noticeable variations with values between 57 and 78
where the value of 67 matches the ratio observed in Orion 
\citep{LangerPenzias1990,LangerPenzias1993}. {\changed We 
therefore assume an error bar of up to 15\,\% for the elemental
abundance ratio.}

The PDR model iteratively computes the steady-state chemistry,
the local cooling and heating processes, and the radiative transfer
to predict the density of the different species and the intensity
of the line and continuum radiation emitted by them.

The restriction of the code to an isotropic illumination is
not a fundamental limitation when comparing different geometries. 
The penetration of an isotropic radiation field can be computed
from the penetration of a uni-directional radiation field by
rescaling the optical depth according to the formalism from
\citet{Flannery1980} and vice versa. \citet{comparison07} showed
a very good match
of the different PDR models when the rescaling was applied. 
When re-interpreting the KOSMA-$\tau$ results for a uni-directional
illumination, one thus has to stretch the width of the outer layers 
relative to the given scale. 

\subsection{{\changed The effect of chemical fractionation on the
\CII{}/\thirteenCII{} ratio}}
\label{sect_planeparallel}

{\changed Paper I already provides some predictions for clump-integrated
intensities for $^{13}$C-bearing species for the special configuration
of a spherical PDRs. To understand in detail how the chemical
fractionation structure can translate into an observable
\CII{}/\thirteenCII{} intensity ratio, we investigate here 
how the depth-dependent carbon fractionation translates into a varying
local \thirteenCII{} and \CII{} emissivity, using the simplified picture
of a plane-parallel configuration described by two parameters
only: the density and the impinging UV field. In the resulting
one-dimensional PDR, different layers reflect a decreasing UV penetration.
This allows us to disentangle the relative contributions of these layers to
the observable \CII{}/\thirteenCII{} intensity ratio. 
The approach ignores all effects resulting from the uncertain details 
of the cloud geometries and of the optical depth of the line
along the line of sight to the observer in a particular geometry.}

\begin{figure}
\includegraphics[width=\columnwidth]{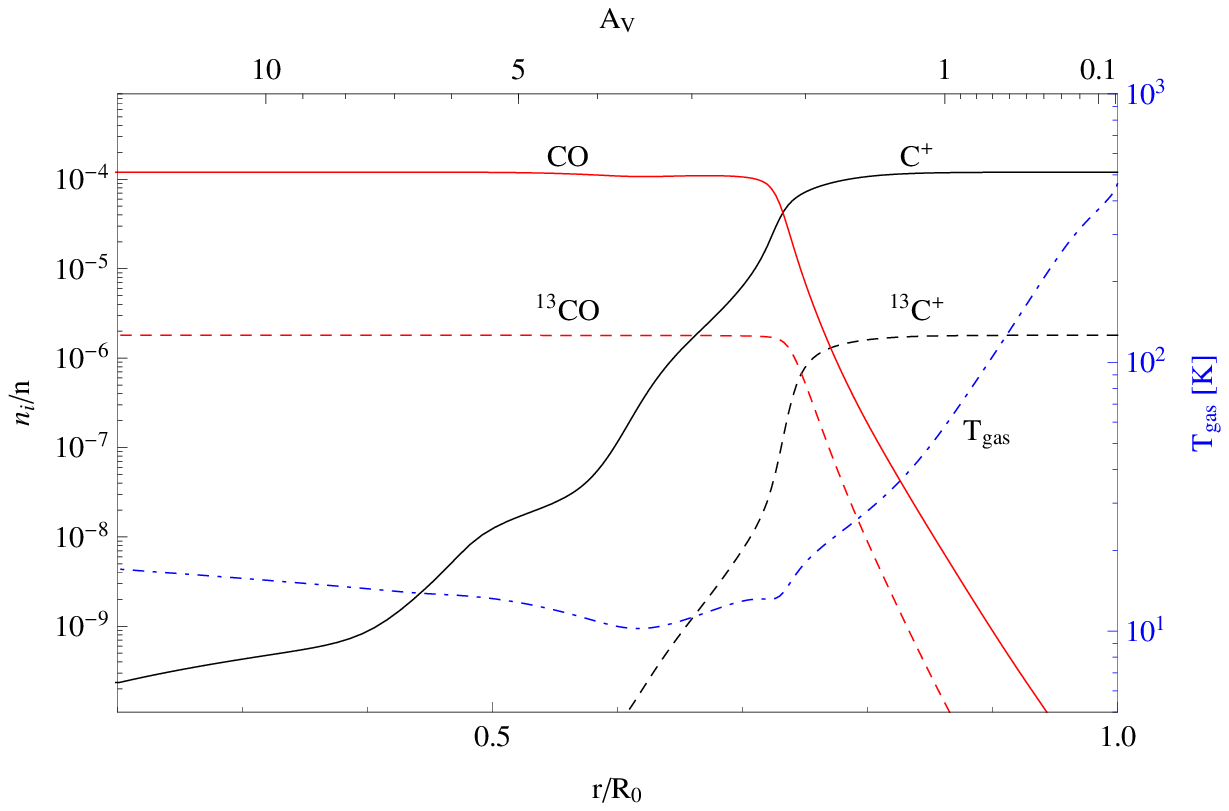}\vspace*{-0.5cm}\\
\includegraphics[width=\columnwidth]{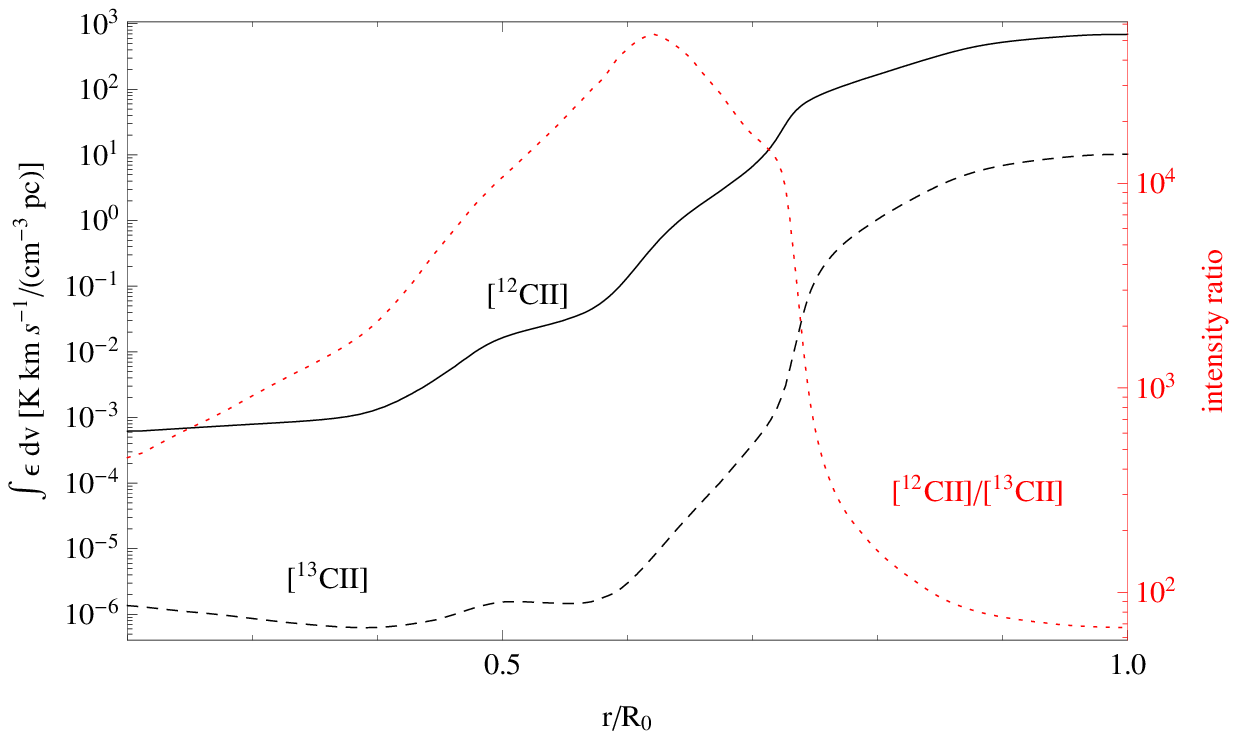}
\caption{{\changed {\bf Top:} Abundance distribution of C$^+$ and CO and the corresponding
$^{13}$C isotopologues for a model with $ n\sub{surf} = 10^4\,\mathrm{cm}^{-3} $, 
$ 1000\,\chi_0 $ and $ 1000\,\mathrm{M}_{\sun}$. The blue curve shows
the corresponding gas temperature structure determining the emissivity.
{\bf Bottom:} Resulting} integrated optically thin line emissivity 
$\int\!\! \epsilon dv$ in K \kms$/($cm$^{-3}$~pc$)$
for the \CII{} and \thirteenCII{} lines (black curves) as a function of the clump
radius. The red line shows the intensity ratio provided by the two curves.}
\label{fig_radialemission}
\end{figure}

{\changed Fig.~\ref{fig_radialemission} shows the temperature and 
abundance structure of the dominating carbon bearing species and the 
resulting \CII{} and \thirteenCII{} line emissivity as a function 
of the relative radius -- or depth from the surface into the cloud, 
respectively --} for an example
model with a density at cloud surface of $10^4$~cm$^{-3}$ and a radiation
field of 1000$\chi_0$. The local \CII{} line emissivity can be computed 
{\changed in terms of a two-level system 
\begin{equation}
\int\!\! \epsilon\, dv =  {h c^3 A \over 8 \pi k \nu^2} \times N_{{\rm C}^+}
	{g_u \exp(-\Delta E/kT\sub{ex}) \over g_l+g_u \exp(-\Delta E/kT\sub{ex})}
\label{eq_emissivity_base}
\end{equation}
with $\Delta E=h \nu = k \times 91.2{\rm K}$, the statistical weights $g_u=4$ and $g_l=2$,
and the Einstein-$A$ coefficient of $A=2.3\times 10^{-6}$~s$^{-1}$ \citep{WieseFuhr2007}, 
providing}
\begin{equation}
\int\!\! \epsilon\, dv \approx 1011 {\rm K\,km s^{-1} \over cm^{-3}~pc} 
	\times N_{{\rm C}^+}
	{2 \exp(-91.2{\rm K}/T\sub{ex}) \over 1+2 \exp(-91.2{\rm K}/T\sub{ex})}
\label{eq_emissivity}
\end{equation}
and we obtain the excitation temperature $T\sub{ex}$ from the
detailed balance including collisional excitation through $H_{\rm 2}$
\citep{FlowerLaunay1977}, atomic hydrogen \citep{LaunayRoueff1977},
and electrons \citep{WilsonBell2002} and radiative de-excitation
and trapping\footnote{$1011\;{\rm K\,km s^{-1}} = 7.11\times 10^{-3}
{\rm erg s^{-1} cm^{-2} sr}$ at the frequency of the \CII{}
transition}. For \thirteenCII{}, this approach sums over all
hyperfine components. The emissivity represents the integrated line intensity
per hydrogen column density without optical depth correction.

{\changed In this example model,} most of the \CII{} emission stems from the cloud surface where
{\changed all carbon is in ionized form and at high temperatures
the C$^+$/$^{13}$C$^+$ fractionation ratio is equal to the elemental isotopic abundance ratio so that
the \CII{}/\thirteenCII{} intensity ratio also matches the elemental ratio.} 
Deep inside the cloud, where {\changed most of carbon is in the form of CO and
\ce{^{13}C+} is only produced by} cosmic ray ionized He$^+$,
the intensity ratio is enhanced by about a factor of 
ten relative to the elemental isotopic ratio, but the C$^+$
{\changed abundance} is reduced
by six orders of magnitude so that the enhanced ratio is hidden
by the surface emission. The region where chemical fractionation
may be detectable is the transition zone from C$^+$ to CO where the
C$^+$/$^{13}$C$^+$ fractionation ratio amounts to a few hundred and the intensity is only a factor 
ten weaker than at the surface (around $r/R_0=0.73$ in 
Fig.~\ref{fig_radialemission}). The actual emissivity ratio
can be boosted to values above the fractionation ratio by radiative excitation
of {\changed C$^+$} in the cool and darker inner regions.
As C$^+$ ``feels'' the strong line emission from the
clump surface, it gains excitation temperatures above the
kinetic temperature while the \thirteenCII{} excitation temperature
closely follows the kinetic temperature. Due to the exponential
term in Eq.~\ref{eq_emissivity} this can provide up to a factor
ten enhancement of the \CII{}/\thirteenCII{} intensity ratio for cold gas at moderate densities 
deeper in the cloud.

\begin{figure}
\includegraphics[width=\columnwidth]{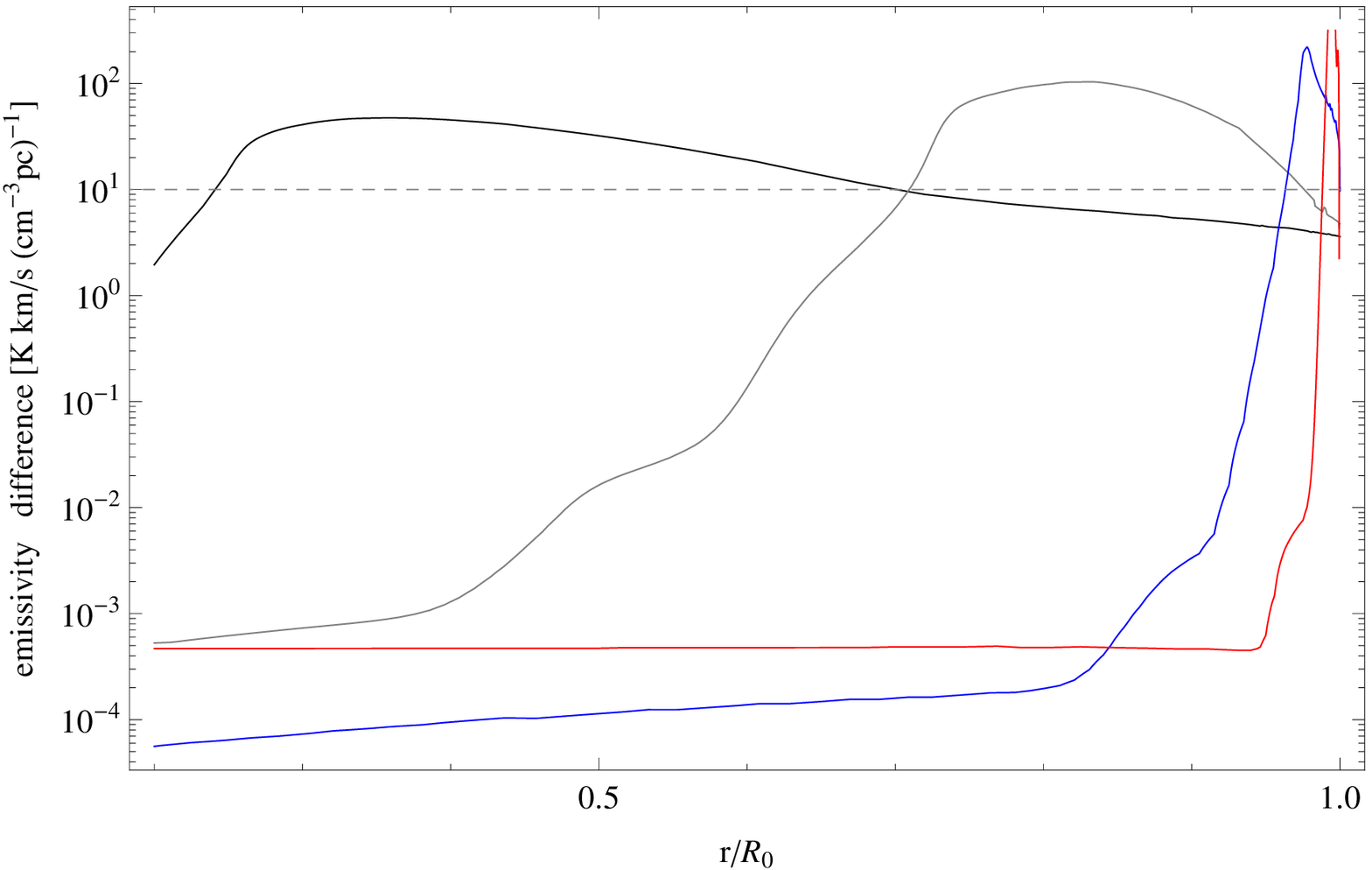}\\
\includegraphics[width=\columnwidth]{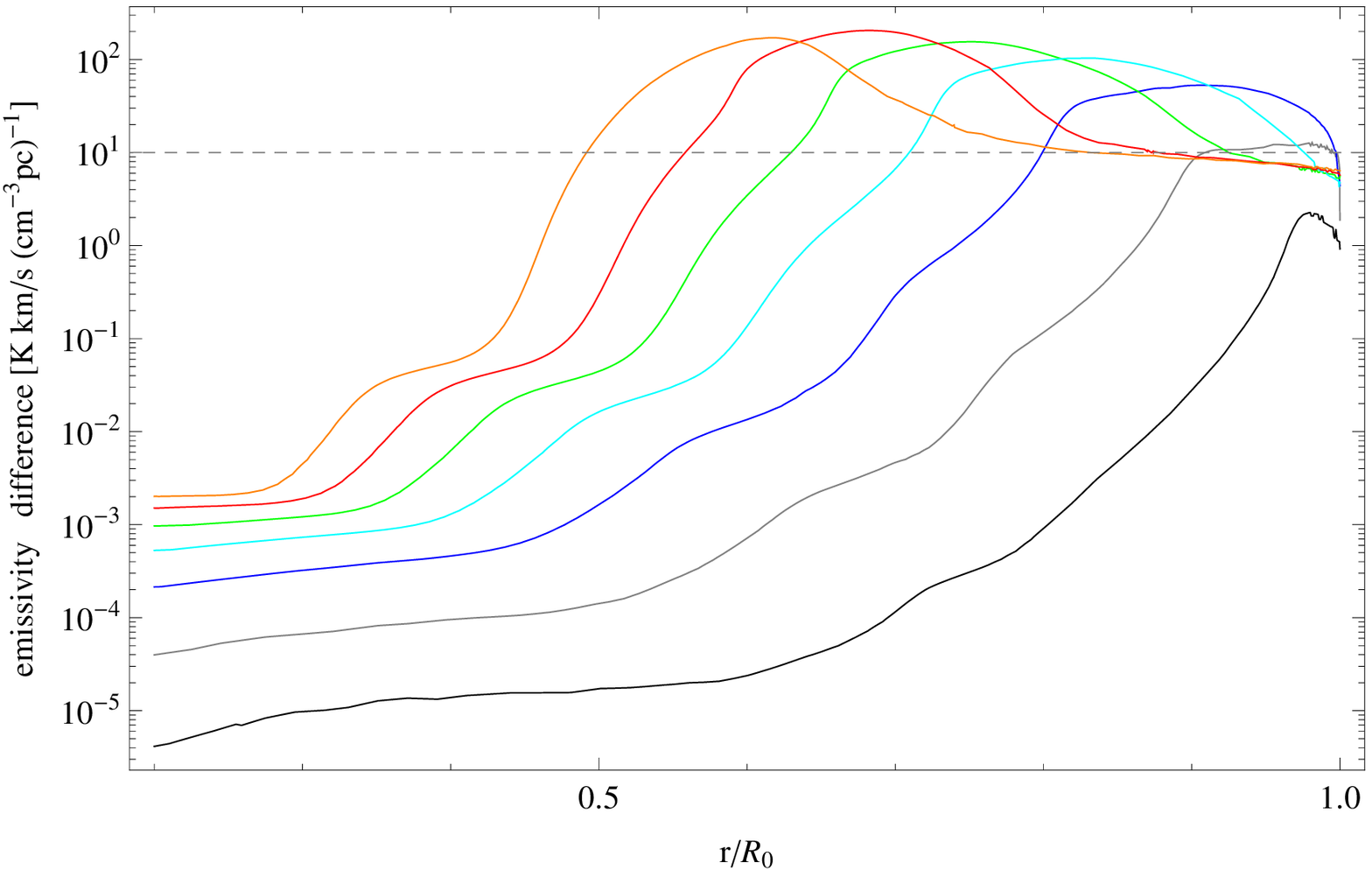}\\
\caption{Differences between \CII{} emissivity and \thirteenCII{} 
emissivity multiplied by the standard elemental isotopic ratio of 67 showing conditions where
a increased C$^+$/$^{13}$C$^+$ fractionation ratio turns into an observable increased 
\CII{}/\thirteenCII{} intensity ratio. The upper
plot uses a radiation field of 1000$\chi_0$ and varies the surface gas
density: $10^3\,\mathrm{cm}^{-3}$ (black), $10^4\,\mathrm{cm}^{-3}$ (grey),
$10^5\,\mathrm{cm}^{-3}$ (blue), $10^6\,\mathrm{cm}^{-3}$ (red).
The lower plot uses a gas density of $n\sub{surf} = 10^4\,\mathrm{cm}^{-3}$
and varies the radiation field: 1$\chi_0$ (black), 10$\chi_0$ (grey), 
100$\chi_0$ (blue), $10^3\chi_0$ (cyan), $10^4\chi_0$ (green), 
$10^5\chi_0$ (red), {\changed $10^6\chi_0$ (orange). The dashed line at}
$10~$K \kms{} (cm$^{-3}$pc)$^{-1}$ {\changed gives a rough indication what differences
might be observable with current technology in a deep integration.}}
\label{fig_radiallineratio}
\end{figure}

To better evaluate for which conditions we might observationally
detect chemical fractionation, we compute the difference between
the \CII{} emissivity and the \thirteenCII{} emissivity scaled
by the elemental abundance ratio. This quantity does not directly reflect the intensity ratio,
{\changed as shown in Fig.~\ref{fig_radialemission}},
but enhancements in this difference indicate conditions
under which an increased C$^+$/$^{13}$C$^+$ fractionation ratio eventually turns into an observable
increased \CII{}/\thirteenCII{} intensity ratio. Figure~\ref{fig_radiallineratio} shows the emissivity
difference as a function of the distance from the outer radius
for varying PDR parameters. In the upper {\changedii panel}, we vary the gas
density for a constant radiation field of 1000$\chi_0$ while the
lower {\changedii panel} shows the impact of different radiation fields for a
constant gas density of $n\sub{surf} = 10^4\,\mathrm{cm}^{-3} $.
Whenever the difference is lower than about 1~K \kms{} (cm$^{-3}$pc)$^{-1}$
no \CII{} enhancement is practically observable, either because the \thirteenCII{}
emission is too weak or because there is no enhanced C$^+$/$^{13}$C$^+$ fractionation ratio.
Three factors determine the detectability of an enhanced fractionation ratio
in the optically thin picture. The radiation field has to be
strong enough to produce a significant column of ionized carbon,
the corresponding layer has to be cool enough so that the
reaction~(\ref{13eq1}) is effective, and the layer of enhanced C$^+$/$^{13}$C$^+$
needs to be thick enough to provide a significant beam filling
in a finite telescope beam. Paper I has shown that these conditions
can be met for a wide variety of conditions, provided that 
there is a sufficient mass of dense gas. With at least $1~M_\odot$
at densities of $10^4$~cm$^{-3}$ a significant fractionation layer
occurs even at high UV fields. 
In the upper {\changedii panel} of Figure~\ref{fig_radiallineratio} we see, 
however, that at densities of $10^5\,\mathrm{cm}^{-3}$ or above, 
the C$^+$--CO transition region, producing an enhanced \CII{}/\thirteenCII{} intensity ratio 
associated with strong emission, is very thin. The lower {\changedii panel} shows
that we find only small \CII{} intensities from clumps exposed
to radiation fields of 10$\chi_0$ or below, making every \CII{}
observation difficult. At all higher radiation fields, fractionation
should be detectable. When considering the full parameter range,
we find some degeneracy in the sense that for higher densities
fractionation may also be observable at low UV fields.
The ideal {\changedii environment} for an increased C$^+$/$^{13}$C$^+$ fractionation
ratio is thus the mildly shielded gas in classical PDRs with 
{\changed $\chi \ga 100 \chi_0$ and} densities in the
order of $10^4-10^5\,\mathrm{cm}^{-3}$. 


\subsection{{\changed Radiative transfer effects on the
\CII{}/\thirteenCII{} ratio}}
\label{sect_clumpymodel}

The \CII{} optical depth can
be computed in the same way as the emissivity (Eq.~\ref{eq_emissivity}) through
\begin{eqnarray}
\int\!\! \tau dv & = & 7.15\times 10^{-18} {\rm km s^{-1} \over cm^{-2}} 
	\times N\sub{C^+} 
	{1- \exp(-91.2{\rm K}/T\sub{ex}) \over 1+2 \exp(-91.2{\rm K}/T\sub{ex})} \nonumber \\
&\approx & 7.15\times 10^{-18} {\rm km s^{-1} \over cm^{-2}} \times N\sub{C^+} 
	{ 32.9 {\rm K} \over T\sub{ex}}\;.
\label{eq_tau}
\end{eqnarray}
For a Gaussian velocity distribution this is equal to 
\begin{equation}
\int\!\! \tau dv  =  {1 \over 2 } \sqrt{\pi \over \ln 2} \Delta v \times \hat{\tau}
\label{eq_tau_gauss}
\end{equation}
where {\changed $\hat{\tau}$ denotes the peak optical depth and $\Delta v$} the FWHM of the velocity distribution.

{\changed Vice versa, we can determine the optical depth from 
a measured \CII{}/\thirteenCII{} intensity ratio (\IR{}), 
at any given frequency in the line when knowing the C$^+$/$^{13}$C$^+$ fractionation ratio (\FR{}),
assuming optically thin \thirteenCII{} lines, and assuming 
equal excitation temperatures of \CII{} and \thirteenCII{}:}
\begin{equation}
{{\it IR}_\nu \over {\it FR}}= {1 - \exp(-\tau_{\rm [C{\mathsc II}]}) \over \tau_{\rm [C {\mathsc II}]}}
\label{eq_optdepth}
\end{equation}
For the line integrated intensity ratio, one has to take the broadening
of the \CII{} line due to the optical depth into account. For a
Gaussian velocity distribution and a line-center optical depth
$\hat{\tau} < 10$, the increase of the line width can be linearly
approximated by the simple factor $1+0.115\hat{\tau}$, and the 
average optical depth measured through the integrated line intensity
is\footnote{The general, more complex expression, also covering
the large optical depth limit, is given e.g. by \citet{Phillips1979}.} 
\begin{equation}
\langle \tau \rangle \approx 0.64 \hat{\tau}\;.
\end{equation}

{\changed To quantify the impact of the line-of-sight optical depth
on the \CII{}/\thirteenCII{} intensity ratio we have to switch from the semi-infinite plane-parallel 
description to a finite geometry, containing a description for
the C$^+$ column density, not only in the direction towards the
illuminating source, but also in the line of sight towards the observer.
This is naturally taken into account when using the
clumpy picture that forms the base of the KOSMA-$\tau$ model,
i.e. considering the case of spherical PDRs. It} introduces the
clump mass, i.e. the absolute radius of the clump, as a
third independent parameter\footnote{We do not include a separate
interclump medium, but rather represent interclump material by the 
equivalent mass of an ensemble of small, transient, low-density clumps.}. 
By integrating over the full volume of each clump
and averaging the intensity over the 
projected area of the spherical model clumps, we mimic the
situation of a fractal cloud, consisting of an ensemble of many clumps,
where individual structures cannot be resolved \citep{Stutzki1998}.

\begin{figure*}
\includegraphics[width=17cm]{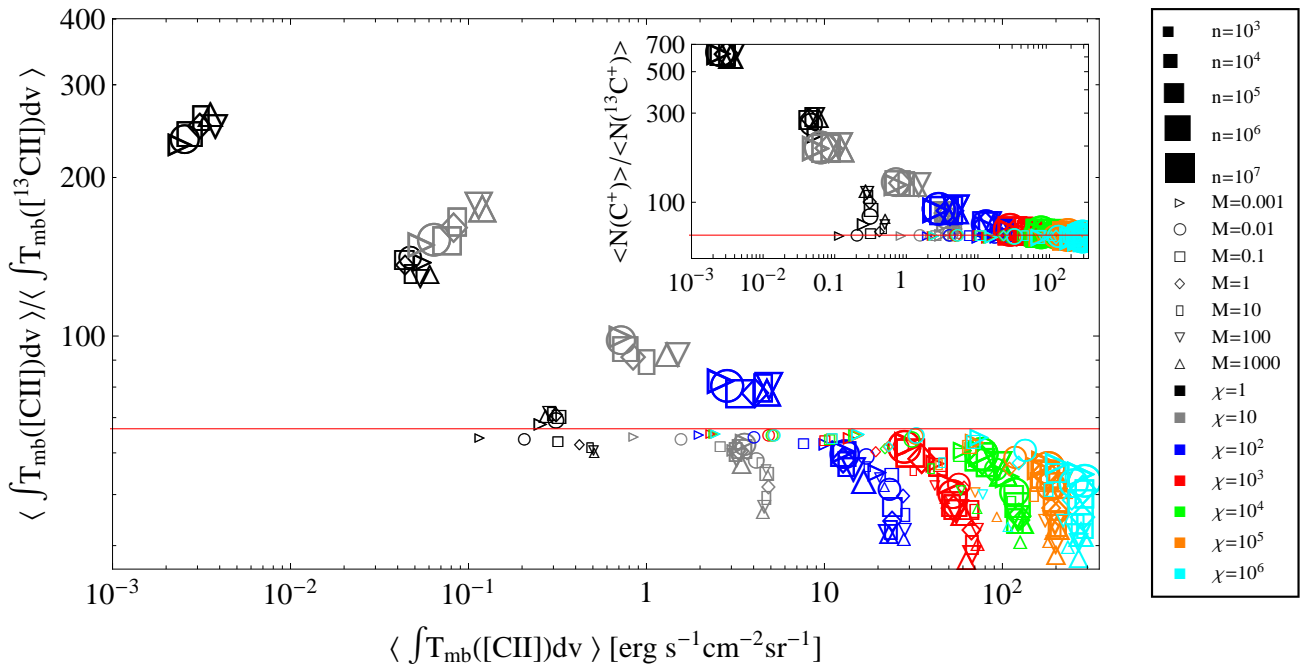}
\caption{Clump integrated intensity ratio $\int I($\CII{}$)dv/\int I($\thirteenCII{}$)dv$ vs.
\CII{} intensity from the KOSMA-$\tau$ model for different parameters.
The model parameters $n$,$M$, and $\chi$, are coded as size, shape, 
and color of the respective symbols. 
The red {\changed horizontal line denotes the assumed elemental abundance ratio
[\ce{^{12}C}]/\ce[\ce{^{13}C}] of 67}.
The insert demonstrates the underlying column density ratio for the different clumps.
For \thirteenCII{}, we sum over all hyperfine components.}
\label{meanTmbc+}
\end{figure*}

In Fig. \ref{meanTmbc+} we compare the ratio of clump 
averaged integrated line intensities, $\int I($\CII{}$)dv/\int I($\thirteenCII{}$)dv$
for a wide range of densities, clump masses and impinging
radiation fields. 
The insert shows the underlying mean column density ratio for the clumps
producing the observed intensity ratio (see paper I).

While the column density ratio falls above the elemental isotopic ratio for all 
model clumps, this does not hold for the \CII{}/\thirteenCII{} intensity ratio. Models 
with low densities and all models with $\chi>100$ show an 
intensity ratio below the elemental ratio, down to a value of 38. {\changed
In these models, the fractionation ratio in the column densities is close to the elemental ratio
so that the reduced intensity ratio is a direct measure of the average optical 
depth of the main isotopic line through Eq.~\ref{eq_optdepth}.
The reduction of the line integrated intensity ratio by a factor two 
corresponds to a moderate \CII{} optical depth,
$\langle \tau \rangle \approx 1.6$ or $\hat{\tau} \approx 2.5$,
respectively, for the individual clumps.}
That means that in all models exposed to high UV fields an ionized
carbon layer of about the same column density is formed, independent
of the other clump parameters. This is naturally explained by
the transition from C$^+$ to CO at about $A_V\approx 0.2$. Different
PDR parameters only change the physical depth and temperature
of this transition point.
Higher \CII{} optical depths only occur if multiple clumps
add up their optical depth on the line of sight at the same velocity
without mutual {\changedii shielding} of the impinging UV field, i.e. in special
geometries.

In the clump-integrated picture, only models with low UV field $\chi$ 
and high densities trace the chemical fractionation, showing an intensity
ratio that is enhanced relative to the elemental abundance ratio. A detailed comparison
shows that even for those clumps some line saturation occurs, so that
the intensity ratio does not measure the true C$^+$/$^{13}$C$^+$ fractionation ratio, but falls
somewhat below it. This result is in contrast to the pure emissivity
considerations in Sect.~\ref{sect_planeparallel} that indicated that
medium-density clumps produce an enhanced intensity ratio.

{\changed
Optical depth effects are always significant in the considered 
geometry with similar column densities of C$^+$ towards the illuminating
sources and towards the observer. C$^+$ is abundant down to a shielding
gas and dust column of $A_V\approx 1$ (see Fig.~\ref{fig_radialemission}).
If all carbon is ionized in this
column, it} provides $N\sub{C^+}\approx 2.5\times 10^{17}$~cm$^{-2}$,
corresponding to a \CII{} optical depth of 0.8 for a slightly subthermal
excitation temperature of $T\sub{ex}=50$~K and the line width of 1.5~\kms{}
used in the model. That means that in isotropic configurations with
narrow lines, as in the clumpy PDR model, \CII{} always turns slightly 
optically thick when the condition for an
efficient fractionation is met. Hence, we find a relatively constant
\CII{} optical depth of 0.8--1.6 for all clumps {\changed that are big enough 
to contain the critical column for $A_V=1$ and that are exposed to a 
radiation field strong enough to ionize the carbon}. 
As a consequence, we expect a behavior as in Fig.~\ref{meanTmbc+} for
isotropic PDRs with narrow velocity distributions.
{\changed Most of the fractionation effect in the C$^+$/$^{13}$C$^+$
abundance ratio that was seen in the local \CII{}/\thirteenCII{} emissivity
ratio in Sect.~\ref{sect_planeparallel}, is therfore hidden from the 
observer by an optical depth close to unity in this case.

A recovery of the true fractionation ratio, i.e. the C$^+$/$^{13}$C$^+$
column densities, through the observable \CII{}/\thirteenCII{} intensity
ratio requires optically thin \CII{} emission, either 
through a much wider velocity dispersion, reducing the \CII{} 
optical depth in a broader line (see Eq.~\ref{eq_tau_gauss}), 
or an anisotropic geometry that provides
a higher column of gas for the UV shielding than in the line-of-sight
towards the observer. Only in these configurations we can expect
to detect chemical fractionation through the \CII{}/\thirteenCII{}
ratio matching the predictions from Sect.~\ref{sect_planeparallel}. 
In contrast we may also find edge-on configurations with a
larger column on the line of sight towards the observer than
towards the illuminating sources (see the Orion Bar in Sect.~\ref{sect_oribar})
where the \CII{} optical depth may grow beyond 2 providing
an intensity ratio much below the local C$^+$/$^{13}$C$^+$ fractionation ratio.

Altogether, we must conclude that in most typical PDRs with bright \CII{}
emission, no significant fractionation is expected to appear in the
C$^+$ column densities, i.e. the fractionation ratio will match the elemental abundance ratio while
the intensity ratio is dominated by the optical depth of the \CII{} line
pushing it below the elemental ratio. In those cases we can only use the intensity ratio
to compute the C$^+$ column density through Eqs.~(\ref{eq_optdepth})
and (\ref{eq_tau}). This is an essential parameter to compare with 
models for deriving e.g. the ionization degree or the photoelectric
heating efficiency. Some carbon fractionation may be detectable through
an increased \CII{}/\thirteenCII{} intensity ratio compared to the
elemental abundance ratio for sources with
high densities and moderate UV fields in case of a favorable geometry
or relatively broad velocity distributions. However, even in these
cases it may be impossible to deduce the exact C$^+$/$^{13}$C$^+$ 
fractionation ratio due to the superimposed optical depth effects.
}

\section{Observations}

\subsection{\thirteenCII{} spectroscopy}
\label{sect_spectroscopy}

{\changed C$^+$} has a single fine structure transition, \CII{} $^2P_{3/2}-^2P_{1/2}$,
at 1900.537~GHz, which for $^{13}$C$^+$ is split into three hyperfine components 
due to the unbalanced additional neutron spin. The frequencies of
these transitions have been determined by \citet{Cooksy1986} based on
the combination of laser magnetic resonance measurements with ab-initio
computations of \citet{SchaeferKlemm1970}.  {\changed Our observations 
from Sect.~\ref{sect_oribar_13cii}} confirm these frequencies with an accuracy of about 
3~MHz, as the relatively narrow \CII{} and \thirteenCII{} lines in 
the Orion Bar match each other and numerous other molecular transitions
to within 0.5~\kms{} in velocity space when using these rest
frequencies. 

We have computed the relative line strengths expected for the 
three hyperfine transitions in local thermodynamic equilibrium (LTE),
which -- for a magnetic dipole transition -- are given by
\begin{equation}
S\sub{hfs} = {(2F'+1)(2F+1)\over 2I+1} \left\{ \begin{array}{ccc} J &  F & I \\
F' & J' & 1\end{array}\right\}^2
\label{eq_spectroscopy}
\end{equation}
(Eq.~8 in \citet{Garstang1995}, normalized to unity). Here, $J' = 3/2$ and 
$J = 1/2$ are the total electronic angular momenta for the upper and lower 
states, $I = 1/2$ is the nuclear spin of the $^{13}$C nucleus, $F'$ and $F$
are the total angular momenta (including nuclear spin) for the upper and 
lower states, and the curly brackets denote a Wigner $6j$-symbol. 
This expression -- which rests on the standard assumptions that magnetic 
dipole (M1) radiation dominates over electric quadrupole (E2) radiation, and 
that the Hamiltonian commutes with all relevant angular momentum operators --
yields a ratio of $0.625 : 0.25 : 0.125$ for the $F' - F = 2 - 1 : 1 - 0 : 1 - 1$
transitions\footnote{Our
line ratios differ from those $(0.444 : 0.356 : 0.20)$ given by \citet{Cooksy1986}
and used in previous astronomical studies. This discrepancy may account for the 
anomaly reported recently by \citet{Graf2012}, who observed a $F = 2 - 1 : F = 1 - 0$
ratio in NGC~2024 that was considerably larger than that predicted by \citet{Cooksy1986}. 
It also requires a minor revision to the elemental isotopic ratio 
inferred by \citet{BoreikoBetz1996} from observations of the $F = 2-1$ and $1 - 0$
transitions observed toward M42. That study applied a correction factor to account 
for the $F = 1 - 1$ transition, which was not covered by the available instrumental 
bandwidth. With our revised line strengths, the correction factor decreases from 
$5/4$ to $8/7$, resulting in a 9.4\% increase in the inferred elemental ratio,
from the range of $52 - 61$ given by \citet{BoreikoBetz1996} to a range of  $57 - 67$. 
This then resolves a small discrepancy, noted by \citet{BoreikoBetz1996}, between the ratio obtained 
from observations of \thirteenCII{} and the value of $67 \pm 3$ inferred from 
observations of $^{13}$CO.}.
All spectroscopic parameters are summarized in Table \ref{tab_spec}.

\begin{table}
\caption{Spectroscopic parameters of the \thirteenCII{} $^2P_{3/2}-^2P_{1/2}$ transition}
\begin{tabular}{rrrrr}
\hline
line & $\nu$ [GHz] & $g\sub{u}$ & $g\sub{l}$ & relative intensity \\
\hline
 $F=2-1$ & 1900.466 & 5 & 3 & 0.625\\
  $F=1-0$ & 1900.950 & 3 & 1 & 0.250\\
  $F=1-1$ & 1900.136 & 3 & 3 & 0.125\\
\hline
\end{tabular}
\label{tab_spec}
\end{table}

\subsection{HIFI measurements}
\label{sect_observations}

We performed \thirteenCII{} observations with the HIFI instrument
\citep{deGraauw} on-board the Herschel satellite \citep{Pilbratt}
towards four bright PDRs in the framework of the key projects
HEXOS \citep{HEXOS} and WADI \citep{WADI}. The first {\changedii region
is the Orion Bar (map center: $5\chr35\cmin20.81\csec$, $-5\deg25'17.1''$),}
 an edge-on, almost linear PDR at a distance
of 415~pc \citep{Menten2007}, formed by material of the Orion Molecular Cloud 1
(OMC-1) exposed to the radiation from the Trapezium cluster providing
an FUV flux of $4-5 \times 10^4 \chi_0$ in terms of the \cite{draine78}
field. 
Previous, spectrally unresolved observations of \CII{} in the
Orion Bar have been carried out by \citet{Stacey1993} and \citet{Herrmann1997}
using the Kuiper Airborne Observatory (KAO). In terms of a pure edge-on 
configuration, the Orion Bar PDR is an improbable candidate for the detection of
{\changed $^{13}$C$^+$} fractionation because of the large column density
of {\changed C$^+$} on the line of sight. However, fractionation effects may
be observable if the \CII{} emission is dominated by the clumpy
structure as seen by \citet{Hogerheijde1995} and \citet{LisSchilke2003}.
{\changedii They found clumps with a density well above $10^6$~cm$^{-3}$
embedded in a more widely distributed ``interclump'' gas with a density
of $10^4-2\times 10^5$~cm$^{-3}$ \citep{Simon1997}.}
Moreover, the gas in front and behind the main Orion Bar should
have a more favorable geometry. By mapping \thirteenCII{}, we
were able to distinguish the three regions.

The second PDR is formed by the molecular interface around
the ultracompact \HII{} region Mon~R2 at a distance of 830~pc
($6\chr07\cmin46.2\csec$, $-6\deg23'08.0''$). Due
to the larger distance, the PDR is weaker in \CII{} than the Orion Bar,
in spite of the larger FUV intensity from the central source of 
about $3\times10^5 \chi_0$ \citep{fuente2010monr2}. The almost
spherical PDR {\changedii geometry} around the \HII{} region provides a
situation close to the plane-parallel model from 
Sect.~\ref{sect_planeparallel} when observing the source center,
with two PDR layers {\changed at high velocities} behind and in front of the \HII{} region
along one line of sight. The velocity dispersion of up to 10~\kms{}
\citep[e.g.][]{Ginard2012} promises a low optical depth at high
integrated line intensities making Mon~R2 one of the best
candidates to detect an enhanced \CII{}/\thirteenCII{} intensity ratio
\citep{Pilleri2012}. The situation is, however,
complicated by the unknown local role of outflows that have been
observed on a somewhat larger scale \citep{Xu2006}.

As a third PDR 
we observed the clump MM2 close to NGC~3603 at a distance of
about 6~kpc \citep{stolte06} ($11\chr15\cmin10.89\csec$, $-61\deg16'15.2''$),
illuminated by a FUV intensity of 
about $10^4 \chi_0$ from  the OB stars in NGC~3603 \citep{Roellig2011}.
The fourth source is the Carina Nebula at a distance of about 2.3~kpc
\citep{Smith2006}, illuminated by the OB stars in the massive clusters
Trumpler 14 and 16 with a FUV flux of about $4 \times 10^3 \chi_0$
\citep{Okada2011}. Here, we observed two interfaces, the 
Carina North PDR ($10\chr43\cmin35.14\csec$, $-59\deg34'04.3''$) close to
Trumpler 14, and the Carina South PDR ($10\chr45\cmin11.54\csec$, $-59\deg47'34.3''$)
south of $\eta$ Car, including the core IRAS-10430-5931.
In NGC3603 and Carina, the PDRs are known to occur at the
surface of clumps, partially visible as pillars in the near infrared.
Therefore, we expect a behavior matching the clumpy model
from Sect.~\ref{sect_clumpymodel} that predicted no observable
enhancement of the \CII{}/\thirteenCII{} intensity ratio for UV fields above $10^2 \chi_0$.

The last PDR is the northern filament in NGC~7023 
($21\chr01\cmin32.4\csec$, $68\deg10'25.0''$) at a distance of
430~pc \citep{vandenAncker1997} illuminated by the pre-main-sequence
B3Ve star HD200775 providing a FUV flux of $1100 \chi_0$
at the position of our observation \citep{Joblin2010,Pilleri2012b}. Here,
previous observations \citep[e.g][]{Gerin1998} show a smooth
density increase towards the filament and a stratified PDR structure
so that we may find the situation modelled in Sect.~\ref{sect_planeparallel}.
The \CII{} mapping results have been presented already by
\citet{Joblin2010}.

All five PDRs were mapped in \CII{} using the on-the-fly (OTF) 
observing mode, however, only the Orion Bar was bright enough 
to detect \thirteenCII{} in the map. Therefore, separate deeper
dual-beam switch (DBS) measurements were
performed towards the given coordinates in the other four PDRs.
Unfortunately, the Orion Bar observations used an LO setting that
placed the $F=1-1$ transition out of the covered IF range, so that 
only the $F=2-1$ and $F=1-0$ components were observed there.

The data were calibrated with the standard HIFI pipeline in the
Herschel Common Software System \citep[HCSS][]{Ott} in version 9.0
(2226). All data presented here are given on the scale of the
antenna temperature corrected for the forward efficiency, 
$T_A^*$. {\changed The preference for this temperature scale is
motivated by the fact that all maps show very extended emission.
The $T_A^*$ scale provides correct brightness temperatures 
if the error beam of the telescope is uniformly filled
by emission of the same magnitude as the main beam. The HIFI
error beam at 1900~GHz is mainly provided by the side lobes of
the illumination pattern appearing at radii of less than
two arcminutes (Jellema et al. in prep). 
Scaling to main-beam temperatures, in contrast, would assume
a negligible contribution of the error beam pickup to the
measured intensity, increasing the resulting temperatures 
by a factor 1.39 \citep{Roelfsema2012}. The two scales
will bracket the true brightness temperature. As the actual
emission decreases with growing distance from the sources, the
$T_A^*$ scale will slightly overestimate the error beam 
contribution and underestimate the brightness temperature, but
due to the very extended emission it should be considerably
better than the main beam temperature scale. We estimate therefore
that the antenna temperature underestimates the brightness 
temperature by 10-20\,\%.}

The resulting spectra suffer from baseline ripples
due to instrumental drifts. They were subtracted using the
{\tt HifiFitFringe} task of the HCSS including the baseline
option with a minimum period of 200~MHz covering all ripples
visible in the spectra. This method attributes
structures wider than 200~MHz to instrumental artifacts and all
narrower structures to sky signal. In this way we
only get reliable intensity information for lines narrower than
30~\kms{}, a condition that is met for all our PDRs. In the
DBS observations, the resulting spectra also suffered from 
self-chopping because the chopper throw of 3$'$ often
ends at OFF positions that are still contaminated by some emission.
This affected the detection of the $F=2-1$ hyperfine component 
in NGC~3603 and Carina falling into the range of self-chopping features
of the main isotopic line. We tried to circumvent this problem by
looking into the individual single-beam switch spectra, that only
use one OFF position left or right of the source, not combining them.
Although they show different degrees of contamination from the
two OFF positions, some contamination always appeared in both
spectra, so that we obtained no qualitative improvement with respect
to a clear detection of the hyperfine component. Therefore, we show
in the following only the dual beam switch spectra providing the
better signal-to-noise ratio.

With the high signal to noise ratio of the \CII{} line in the
Orion Bar map we noticed a small zig-zag structure in reported
positions and the resulting Orion Bar map, originating from the
alternating direction in the observation of subsequent OTF lines.
An ad-hoc correction of the pointing information by 1.4$''$, {\changed
mutually shifting the OTF lines relative to each other,} resulted in a 
straight Orion Bar structure. This correction will be implemented
in future versions of the HIFI pipeline, but for this paper
the correction just gives an estimate for the pointing accuracy.
The original Orion Bar map also suffered from relatively strong
emission at the OFF position ($5\chr35\cmin44.92\csec$, $-5\deg25'17.1''$)
visible as absorption feature in numerous points of the map. A
separate observation of the OFF position was performed using the
internal HIFI cold load as primary reference and a secondary
OFF position more than 12$'$ away from the Bar ($5\chr35\cmin55.0\csec$, 
$-5\deg13'18.1''$) as secondary reference. Fortunately, it turned out
that the baseline from the subtraction of the cold load spectrum was good
enough so that the secondary OFF position did not have to be used
because we detected even from this very remote position \CII{}
emission as bright as 10~K (see Fig.~\ref{fig_kao}) in agreement
with the large-scale map of \citet{Mookerjea2003}.
This has to be considered as a very widely distributed emission floor
in the Orion region.

\section{Observational results}\label{sec-results}

\begin{table*}
\caption{Summary of integrated line intensities from the observations {\changed and derived parameters.}}
\begin{tabular}{lrrrrrrrr}
\hline
source & \hspace*{-1cm}integration range & \CII & \thirteenCII{} $F=2-1$ & $F=1-0$ & $F=1-1$ & \IR{\tablefootmark{a}}
& $\langle \tau\sub{[C{\sc II}]} \rangle $ &  $N\sub{C^+}$\tablefootmark{b} \\
& [\kms{}] & [K \kms{}] & [K \kms{}] & [K \kms{}] & [K \kms{}] & & & \hspace*{-0.5cm}[$10^{18}$~cm$^{-2}$] \\
\hline
Orion Bar, peak &
$ 7-13 $ & $ 857 \pm 5 $ & $ 17.2 \pm 1.0 $ & $ 7.8 \pm 0.5 $ & $ - $ &  $ 30 \pm 2$ & $ 1.9 \pm 0.2$  & $10.1$\\
Orion Bar, ridge\tablefootmark{c} &
$ 7-13 $ & $ 772 \pm 5 $ & $ 12.6 \pm 1.0 $ & $ 5.1 \pm 0.5 $ & $ - $ & $ 38 \pm 3$ & $ 1.3 \pm 0.2$ & $ 7.2$ \\
Orion Bar, front\tablefootmark{d} &
$ 7-13 $ & $ 506 \pm 7 $ & $ 4.3 \pm 0.7 $ & $ 2.1 \pm 0.6 $ & $ - $ & $ 69 \pm 12$ & $ 0.0 \pm 0.4$  & $2.3$\\
Orion Bar, back\tablefootmark{e} &
$ 7-13 $ & $ 529 \pm 6 $ & $ 6.3 \pm 0.6 $ & $ 2.9 \pm 0.6 $ & $ - $ & $ 50 \pm 6$ & $ 0.7 \pm 0.3$  & $3.9$\\

Mon~R2, total & $ 5-25 $ & $ 362 \pm 5 $ & $ - $ & $ 2.9 \pm 0.7 $ & $ 0.7 \pm 0.5 $  & $ 38 \pm 10$ & $ 1.3 \pm 0.6$ &  $5.7$\\
Mon~R2, blue & $ 5-12.5 $ & $ 173 \pm 3 $ & $ - $ & $ 2.4 \pm 0.7 $ & $ 0.8 \pm 0.3 $  & $ 20 \pm 5$ & $ 3.2 \pm 0.8$ &  $4.8$\\
Mon~R2, red & $ 12.5-25 $ & $ 188 \pm 4 $ & $ - $ & $ 0.7 \pm 0.5 $ & $ -0.3 \pm 0.4 $  & $ 170 \pm 120$ & $ 0.0 \pm 0.5$ & $0.9$\\
NGC~3603 & $ 10-19$  & $ 130 \pm 2 $ & $ - $ & $ 1.4 \pm 0.2 $ & $ 0.55 \pm 0.15 $ & $ 25 \pm 5 $ & $2.4 \pm 0.4$  & $3.6$\\
Carina N & $ -20 - -5 $ & $ 143 \pm 4 $ & $ - $ & $ 1.3 \pm 0.3 $ & $ 0.7 \pm 0.3 $ & $ 27 \pm 7 $ & $2.2 \pm 0.6$ & $5.0$\\
Carina S & $ -38 - -25 $ & $ 38 \pm 2 $ & $ 2.7 \pm 0.3 $\tablefootmark{f} & $ -0.1 \pm 0.3 $ & $ 0.0 \pm 0.3$ & $ > 22$ & $<2.9$ 
	& $0.2$\tablefootmark{g}\\
NGC~7023 & $ -1 - 7 $ & $91 \pm 2$ & $1.07 \pm 0.05$ & $0.53 \pm 0.07$ & $ 0.19 \pm 0.04$ & $51 \pm 6$ & $0.6 \pm 0.3$ & $1.0$\\
\hline
\end{tabular}
\tablefoot{
\tablefoottext{a}{\CII{}/\thirteenCII{} intensity ratio. From the sum over all \thirteenCII{} hyperfine components 
assuming the ratios from Table~\ref{tab_spec} to correct for non-detected components.}
\tablefoottext{b}{Assuming a uniform Gaussian velocity profile for C$^+$ and $^{13}$C$^+$.}
\tablefoottext{c}{Average over the {\changedii $\approx 1000$ square-arcsec} with $I($\CII{}$) \ge 700$~K~\kms{} representing the main Bar emission {\changedii (see contours in Fig.~\ref{fig_bar12cii}}.}
\tablefoottext{d}{Average over the {\changedii $\approx 750$ square-arcsec} with $450$~K~\kms{}$ \le I($\CII{}$) \le 650$~K~\kms{} north-west of the Bar.}
\tablefoottext{e}{Average over the {\changedii $\approx 900$ square-arcssec} with $450$~K~\kms{}$ \le I($\CII{}$) \le 650$~K~\kms{} south-east of the Bar.}
\tablefoottext{f}{The identification of this emission as \thirteenCII{} $F=2-1$
is questionable because of the small separation from the \CII{} line. It
may rather represent a different velocity component.}
\tablefoottext{g}{Assuming $\tau=0.0$ to compute the column density.}
}
\label{tab_intensities}
\end{table*}

Table~\ref{tab_intensities} lists the integrated intensities of the \CII{}
line and the \thirteenCII{} components for all sources. The relative 
intensities of the \thirteenCII{} hyperfine components are roughly
consistent with the intensity ratios from Table~\ref{tab_spec}. However,
in the observations with the best signal-to-noise ratio, the Orion Bar peak
and NGC~7023, we find a somewhat lower $F=2-1/F=1-0$ ratio of $2-2.4$
instead of $2.5$. The uncertainty
of the integrated intensities consist of three contributions: radiometric noise, 
the definition of the appropriate velocity integration range, and the uncertainty in 
the baseline subtraction. When measuring the noise in the spectra, we found
that its contribution to the total uncertainty is actually negligible
compared to the two other sources of uncertainty for all our observations.

The definition of the best integration range asks for a compromise because
broad wings in the \CII{} profile, in particular in Mon~R2, favor
very broad integration ranges {\changed to trace the whole \CII{} emission}
while the small separation between the \CII{}
line and the \thirteenCII{} $F=2-1$ transition and wavy baseline structures
ask for narrow integration ranges. As the wings are not detectable in the
\thirteenCII{} components and their intensity ratio forms the main focus of this study,
we used common relatively narrow integration ranges here (column 2 of 
Tab.~\ref{tab_intensities}). In this way, we ignore the wing emission that
is detectable in the main isotope line, but compare the same velocity ranges
for the isotopic ratios. In the extreme case of Mon~R2, the integrated \CII{}
intensity is higher by 15~K \kms{}, i.e. 4\,\%, compared to the value in Tab.~\ref{tab_intensities}
when using a very broad integration range including all the wings and in
the Orion Bar the value would be increased by 7~K \kms{}, i.e. 1\,\%, in this way.
For the other sources no wing components can be detected due to 
self-chopping signatures in the spectra (see Fig.~\ref{fig_3603_carina}).

The uncertainty that actually determines the error bars given in
Tab.~\ref{tab_intensities} stems from the baseline subtraction in the
spectra. {\changed Only to a small degree they are given by radiometric
noise, but they are dominated by correlated residuals in the baselines
that are not easily described by statistic measures.}
We performed numerical experiments by varying the number of 
sinusoidal components in the {\tt HifiFitFringe} task between 1 and 4
and changing the size of the window that is masked, i.e. excluded
from the baseline fit, from the pure integration range to a range
that is about three times wider in five steps. We inspected all baseline
fits manually excluding those where self-chopping signatures or line
wings affected the fit. The uncertainties in Tab.~\ref{tab_intensities}
describe the variation in the remaining sample in terms of the total covered
range, i.e. they do not quantify $1 \sigma$ errors but the extremes of that 
sample. It is obvious that the relative errors are small for the Orion Bar
with narrow lines while they increase with the line width and the
resulting larger baseline uncertainty.  For the figures shown in the next 
section we used the subjective ``best baseline fit'' from the sample.

\subsection{Orion Bar}
\label{sect_oribar}

\subsubsection{\CII{}}
\label{sect_oribar_cii}

\begin{figure}
\resizebox{\hsize}{!}{\includegraphics[angle=90]{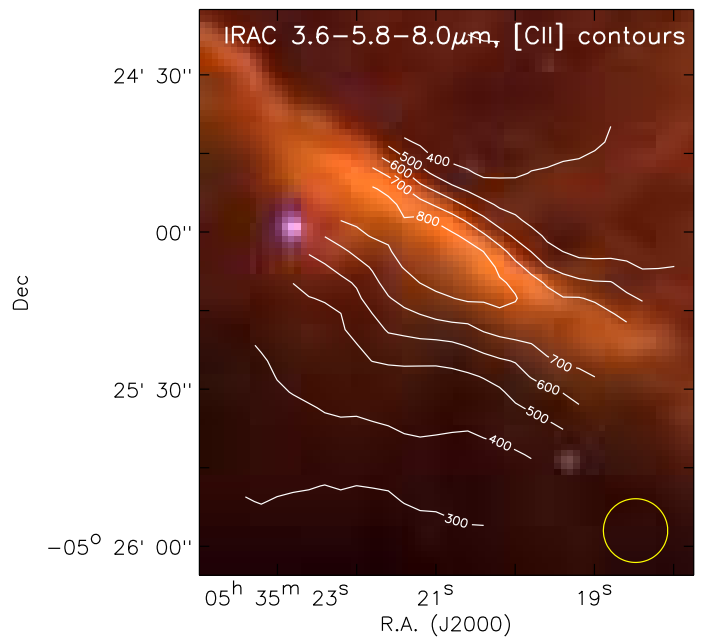}}
\caption{Integrated \CII{} contours over-laid on the false-color
IRAC map  of the Orion Bar {\changed (red: 8.0~$\mu$m: 0--3000~MJy/sr,
green: 5.8~$\mu$m: 0--7000~MJy/sr, blue: 3.6~$\mu$m: 0--10000~MJy/sr)}.
\CII{} is mapped in a strip perpendicular
to the Orion Bar. Intensities are labelled in units of K \kms. The
colors in the IRAC map are provided by the signal in the 3.6, 5.8,
and 8 $\mu{}$m channels. The yellow circle indicates the HIFI beam size.}
\label{fig_bar12cii}
\end{figure}

\CII{} was mapped along a strip perpendicular to the direction of 
the Bar. Figure~\ref{fig_bar12cii} shows the overlay of the integrated intensity
on the corresponding IRAC 8\,$\mu$m image. \CII{} shows a
very smooth structure without any indication for  clumpiness
 as seen e.g. in the different isotopologues of CO,
HCN, HCO$^+$, CS and H$_2$CO by \citet{Hogerheijde1995} and \citet{LisSchilke2003}
{\changedii and to a lower degree also in the C91$\alpha$ carbon radio 
recombination line by \citet{Wyrowski1997}}.
This indicates that \CII{} traces mainly the widely distributed,
relatively uniform interclump medium. 
{\changedii Position, structure, and overall extent of the \CII{} Orion
Bar emission are similar to the C91$\alpha$ radio recombination line map. 
That map shows, however, a larger contrast and some clumpy substructure
within the emission peak. To some degree that may be due to insufficient
short-spacing of the interferometer data, but it probably also reflects
real differences in the excitation of the C$^+$. As the C91$\alpha$
emission depends much more sensitively on the gas density and temperature
than \CII{} \citep[see e.g.][]{Natta1994}, the ratio of the two lines may
provide a way to better characterize the transition between the widely
distributed gas and the embedded clumps.} The \CII{} emission peaks
approximately 10$''$ deeper in the cloud compared to the PAH emission
traced by the IRAC 8\,$\mu$m map. This coincides approximately with the peak of
the H$_2$ 1-0 S(1) line measured by \citet{Walmsley2000} and correlates
very well with the C$_2$H emission observed by \citet{vdWiel2009}. The detailed
comparison of the stratification in the different lines will be the topic
of a forthcoming paper (Makai et al. in prep).

\begin{figure}
\resizebox{\hsize}{!}{\includegraphics[angle=90]{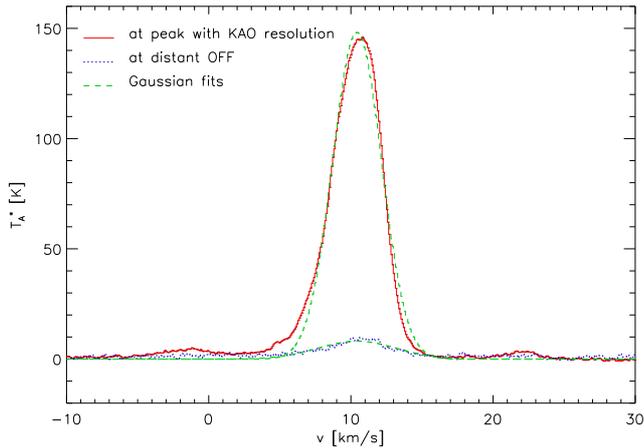}}
\caption{\CII{} profile towards the peak of the {\changedii Orion Bar}
emission {\changedii ($5\chr35\cmin20.6\csec$, $-5\deg25'5''$)} averaged over
a 43$''$ (FWHM) Gaussian beam representing the resolution of the 
previous KAO observations (solid line). The dotted spectrum
represents the widely distributed emission as measured on our
secondary OFF position 12.3$'$ away from the Orion Bar in a region 
without molecular emission. {\changed The dashed lines show Gaussian fits
to the two profiles.}}
\label{fig_kao}
\end{figure}

We can compare our data with previous KAO observations of \CII{} in Orion.
\citet{Boreiko1988} and \citet{BoreikoBetz1996} obtained frequency-resolved
spectra towards $\Theta^1$C, but not towards the Orion Bar.  The spectrally
unresolved maps by \citet{Stacey1993} and \citet{Herrmann1997} showed 
approximately the same integrated intensity at the peak of the Orion Bar and towards
$\Theta^1$C. The $\Theta^1$C value from \citet{BoreikoBetz1996} of
$3.8\times 10^{-3}$ erg cm$^{-2}$ s$^{-1}$ sr$^{-1}$ is, however, 16\,\%
lower than the $4.5\times 10^{-3}$ erg cm$^{-2}$ s$^{-1}$ sr$^{-1}$ 
derived for the integrated \CII{} intensity at the Orion Bar by \citet{Herrmann1997}
scaling all intensities to match the previous value
from \citet{Stacey1993} towards Orion KL.
 Figure~\ref{fig_kao} shows our \CII{} HIFI profile 
towards the peak in the Orion Bar
after convolution to the 43$''$ resolution of the KAO. The integrated intensity
of 585~K\;\kms{}$ = 4.1\times 10^{-3}$~erg~cm$^{-2}$~s$^{-1}$~sr$^{-1}$
falls between the two quoted KAO values. The discrepancy {\changedii with} the lower value 
from \citet{BoreikoBetz1996} can be partially explained by self-chopping
of the KAO observations in the very extended emission. Our observation
of the secondary OFF position (also shown in Fig.~\ref{fig_kao}) gives
an estimate for this extended emission providing a 6\,\% correction
of the KAO values. The 9\,\% discrepancy {\changedii with} the \citet{Herrmann1997}
value could be due to the uncertainty in the error beam pickup discussed
above or the scaling method used in their paper.

\begin{figure}
\resizebox{\hsize}{!}{\includegraphics[angle=90]{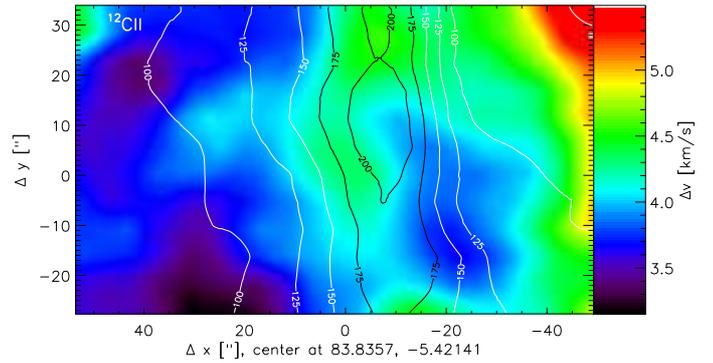}}
\caption{Contours of the \CII{} line peak intensity overlaid on the line width
(FWHM) obtained from a Gaussian fit to the \CII{} lines at all points in the 
Orion Bar map.} 
\label{fig_linewidth}
\end{figure}

As {\changed we find almost perfect Gaussian profiles for the lines 
in the Orion Bar (see Fig.~\ref{fig_kao})} we performed Gaussian
fits to all \CII{} spectra.  In Fig.~\ref{fig_linewidth} we show the 
resulting map of amplitudes and widths. As the 
observations by \citet{BoreikoBetz1996} indicated
partially optically thick lines, we expect that the lines become broader
towards the peak of the emission due to optical depth broadening. The
map, however, does not show any significant change of the line width across the
Orion Bar. The FWHM of the line falls between 4.0 and 4.5~\kms{} only
turning broader for the more diffuse emission in the north. Consequently,
the amplitude map gives a perfect match to the integrated line map in
Fig.~\ref{fig_bar12cii}. As we confirm the significant optical depth
in the following section, this indicates that the \CII{} line must
be composed of many velocity components which are individually optically
thick, instead of a continuous microturbulent medium that only turns
optically thick at the largest column density sum.
The corresponding map of line velocities only shows a small gradient
along the Bar with velocities of about 11~\kms{} in the north-east and
10.5~\kms{} in the south-west.

\subsubsection{\thirteenCII{}}
\label{sect_oribar_13cii}

\begin{figure}
\resizebox{\hsize}{!}{\includegraphics[angle=90]{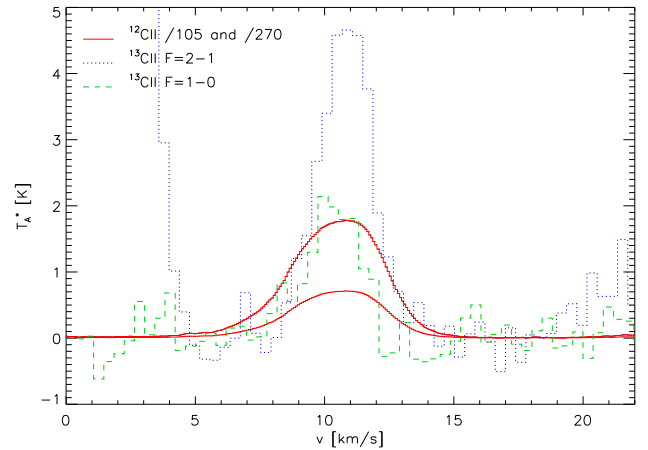}}
\caption{Average profile of the two strongest \thirteenCII{} hyper-fine
components in the Orion Bar compared to the \CII{} profile scaled by 
factors of 0.625/67 and 0.25/67. The profiles were averaged over all 
pixels with a \CII{} integrated intensity above 700K km/s.}
\label{fig_oribar_ratios}
\end{figure}

Due to the narrow line width in the Orion Bar, we are able
to detect the $F=2-1$ and $F=1-0$ lines of \thirteenCII{} without
noticeable blending with the \CII{} line. Figure~\ref{fig_oribar_ratios}
shows the line profiles of the two \thirteenCII{} components when averaged over
the region with the brightest emission, i.e. with \CII{} intensities
of more than 700~K~\kms. For a comparison we have added the corresponding
\CII{} spectra, scaled by factor of $0.625/67$ and $0.25/67$. This scaling 
corresponds approximately to the line intensity of the two \thirteenCII{} 
transitions that would be expected when we assume the canonical abundance ratio 
and optically thin lines.

\begin{figure}
\resizebox{\hsize}{!}{\includegraphics[angle=90]{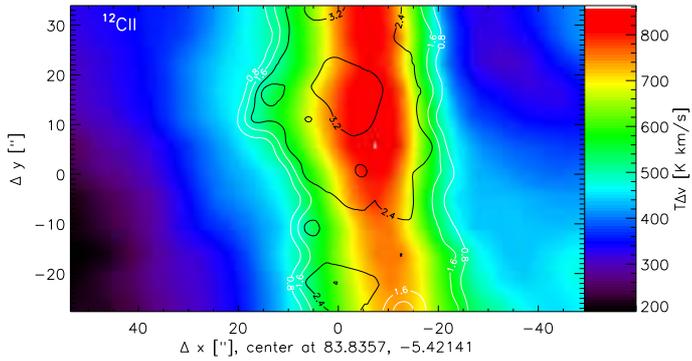}}
\caption{Contours of the \CII{} line center optical depth, 
derived from the \CII{}/\thirteenCII{} ratio, overlaid on
the integrated \CII{} intensities observed towards the Orion Bar.} 
\label{fig_map_tau}
\end{figure}

When considering the line amplitudes, we find that both hyperfine transitions
are approximately 2.5 times brighter than expected from optically thin \CII{}
emission and the normal isotopic ratio. Assuming the canonical abundance
ratio, this corresponds to a line-center optical depth, $\hat{\tau}$ of 2.2. 
The good signal to noise ratio of the data allows us to perform Gaussian fits 
to the hyperfine lines over the full map providing a full map of intensity 
ratios. Using Eq.~\ref{eq_optdepth} and assuming no chemical fractionation,
we can translate this into a map of optical depths.
Figure~\ref{fig_map_tau} shows the resulting line center optical depth, $\hat{\tau}$,
overlaid on the integrated \CII{} intensities. Here, we have added both
hyperfine transitions to increase the signal to noise.
No reliable values can be derived for $\tau\la 0.8$ as Eq.~\ref{eq_optdepth}
turns very flat for low optical depths. Hence, all points with lower optical
depth have been masked-out. We find a good correlation of the
optical depth structure with the intensity map. The map confirms that 
the area of bright \CII{} emission has optical depths above two and the
peak optical depth of 3.4 occurs close to the intensity peak, but we also
see a small general shift of the optical depth structure relative to the
intensity structure away from the exciting Trapezium cluster. This is
consistent with the picture of hotter gas at the PDR front and somewhat
cooler gas, but a higher column density deeper into the cloud.

\begin{figure}
\resizebox{\hsize}{!}{\includegraphics[angle=90]{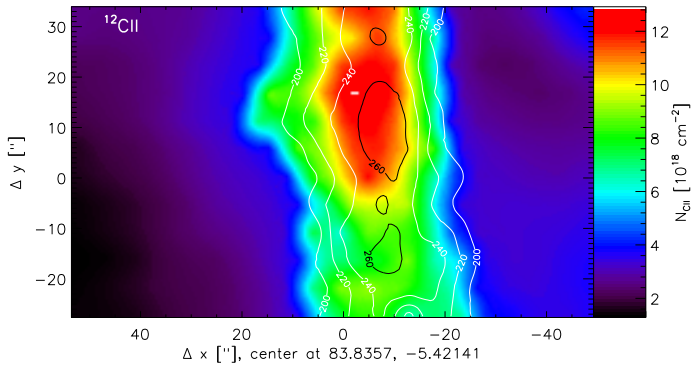}}
\caption{Contours of the \CII{} excitation temperature overlaid on
the derived column densities towards the Orion Bar.} 
\label{fig_map_tex_ncol}
\end{figure}

To better quantify this, we use Eq.~\ref{eq_tau} to compute the
excitation temperatures, and Eq.~\ref{eq_emissivity} together with the
optical depth correction in Eq.~\ref{eq_optdepth} to compute the
total C$^+$ column density. The resulting distribution is shown in
Fig.~\ref{fig_map_tex_ncol}.\footnote{When scaling the measured intensities
to the main beam efficiency, i.e. assuming less extended emission,
the optical depth does not change, but the excitation temperature
and C$^+$ column densities increase by 30\,\% relative to the values
given here.} In the range of significant optical depths,
the excitation temperatures fall between 190 and 265~K. For
$\tau \le 0.8$ we assumed a constant excitation temperature of 190~K
and directly used the optically thin approximation of Eq.~\ref{eq_emissivity}
to compute the C$^+$ column density.

Our excitation temperatures are in agreement with the average
value of {\changedii 220~K and 215~K derived by \citet{Herrmann1997} 
and \citet{SorochenkoTsivilev2000}, respectively, for the Orion Bar.
Interclump gas at this temperature with a density around $10^5$~cm$^{-3}$ 
also explains the C91$\alpha$ observations by \citet{Wyrowski1997}.
However, \citet{Natta1994} and \citet{Walmsley2000} obtain much
higher temperatures from the radio recombination lines. This could indicate
nonthermal emission, dielectronic recombination, or simply
different gas components contributing to the \CII{} and C91$\alpha$
emission \citep{Natta1994,SorochenkoTsivilev2000}.}

At the scale of the KAO resolution, our approach provides a
column density of $7-8\times 10^{18}$~cm$^{-2}$ being more
than two times higher than the value of \citet{Herrmann1997}, 
consistent with our \CII{} optical depth exceeding the value
derived by \citet{BoreikoBetz1996} by about the same factor.
The discrepancy can be explained to a large degree by their
use of the line-averaged optical depth, underestimating the
peak depth by a factor 0.64 for a Gaussian velocity distribution
(see Sect.~\ref{sect_clumpymodel}), and their use of 
different spectroscopic parameters, as discussed in 
Sect.~\ref{sect_spectroscopy}. Our peak C$^+$ column density of 
$13\times 10^{18}$~cm$^{-2}$ implies a configuration where
most of the line-of-sight column density of the Orion Bar of
$N\sub{H} \approx 13\times 10^{22}$~cm$^{-2}$ \citep[e.g.][]{Hogerheijde1995}
forms a PDR with complete ionization of the gas phase carbon
-- {\changed abundance $X($C$)=n($C$)/(n($H$)+2 n($H$_2))=1.2\times 10^{-4}$ 
{\changedii for dense clouds} \citep{wakelam08}}.
A surprising result is, however, the relatively large
C$^+$ column density in front and behind the Bar. Molecular line 
observations \citep[e.g.][]{Hogerheijde1995, vanderWerf1996} deduced
column densities that are a factor ten lower than in the Bar.
Consequently, at least half of the \CII{} emission in the
veil must stem from atomic, more diffuse material.

Fig.~\ref{fig_map_tex_ncol} also shows the relative shift between
the column density and the excitation temperature distributions
by 5--8$''$ with the higher temperatures towards the illuminating source.
The new observations thus resolve a temperature gradient in C$^+$
consistent with the overall stratified structure of the Orion Bar PDR.
The resulting intensity profile then represents a convolution of 
density and temperature structures.

\subsubsection{Line averages}

To verify the dependence of the results on the assumption of the
Gaussian velocity distribution, we repeated the analysis for selected
subregions when considering the line-averaged \CII{}/\thirteenCII{}
intensity ratio, i.e. $\langle \tau \rangle$
instead of $\hat{\tau}$. This is also the only possible analysis for the
other sources where the signal-to-noise ratio is lower and the line 
profiles are highly structured so that the assumption of a Gaussian 
velocity distribution is clearly violated. {\changed The use of the
line averaged optical depths, $\langle \tau \rangle$, corresponds to the
assumption of a macro-turbulent situation, where \CII{} turns
independently optically thick in all velocity channels, while
the $\hat{\tau}$ method assumes a microturbulent Gaussian velocity distribution
for all tracers, where \CII{} turns optically thick only in the line center.
A preference for one of the methods is only possible based on detailed 
knowlegde of the source structure.
The $\langle \tau \rangle$ results for all sources} are included in the
last two columns in Table~\ref{tab_intensities}.

For the Orion Bar, the $\langle \tau \rangle$ results always fall 
below the column densities from the line-center optical depth by
30--35\,\% in spite of the almost Gaussian line profiles. A possible
cause is the systematic difference in the line width between the 
\thirteenCII{} and the \CII{} lines going beyond the expected
optical-depth broadening. As seen in the average spectrum shown 
in Fig.~\ref{fig_oribar_ratios}, but also in the map of \thirteenCII{}
line widths corresponding to Fig.~\ref{fig_linewidth}, the \thirteenCII{}
line is always 1.0--1.5~\kms{} narrower than the \CII{} line. 
As a consequence, the intensity ratio is even larger than the standard elemental isotopic ratio in the 
line wings (well visible in the blue wing in Fig.~\ref{fig_oribar_ratios}).
Together with the constant \CII{} line width discussed in the previous
section, the enhanced intensity ratio in the wings indicates different velocity
distributions for C$^+$ and $^{13}$C$^+$ so that neither the $\hat{\tau}$
method nor the $\langle \tau \rangle$ method can provide accurate results.
The 30--35\,\% of the emission remaining in the wings of the \CII{}
line when fitting it with the profile of the \thirteenCII{} lines
provide an estimate for the accuracy of the method.

The $\langle \tau \rangle$ values from the average spectra in
Table~\ref{tab_intensities} also confirm the spatial gradient of the 
column densities discussed for Fig.~\ref{fig_map_tex_ncol}.
When analyzing the bright gas in front and behind the Orion Bar by 
selecting all pixels with a \CII{} intensity between 450 and 650~K~\kms
north-west, i.e. in the direction of the illuminating Trapezium stars, and
south-east of the Bar, we find significantly higher optical depths
behind the Bar while the gas in front of the Bar is {\changed
typically hotter by about 20~K, but thinner by a factor 1.7.}

None of the average spectra show an indication for an enhanced
C$^+$/$^{13}$C$^+$ fractionation ratio. This also applies to the spectra from the veil in front
and behind the Bar where we were not able to detect a significant
deviation from the elemental abundance ratio due to the relatively larger baseline 
uncertainties at lower intensities of the \thirteenCII{} lines.
Our \thirteenCII{} observations are consistent with negligible 
{\changed carbon} fractionation and an enhanced optical depth of the main isotopic 
line for high column densities. This is in agreement with a previous 
detection of the \thirteenCI{} $F=5/2-3/2$
transition by Keene et al. (1998) in a selected clump where
the observed $^{13}$C/C ratio showed no fractionation effect in contrast
to the complementary observations for $^{13}$C$^{18}$O and C$^{18}$O. 

An indication for an enhanced fractionation ratio is, however, seen in the
velocity structure when considering the wings of the bright line 
profiles. The broader \CII{} lines showing enhanced \CII{}/\thirteenCII{} intensity ratio in the 
wings (e.g. in Fig.~\ref{fig_oribar_ratios}) indicate that an additional
gas component, with broader velocity dispersion and enhanced fractionation ratio,
thus invisible in \thirteenCII{}, contributes to the \CII{} profiles.
The constancy of the \CII{} line width then could be explained by
the mutual compensation of the relatively larger contribution from 
the low-velocity dispersion Orion Bar material with the optical depth 
broadening there. To sustain the enhanced intensity ratio, the broad velocity
gas component should have moderate gas densities of $10^3-10^4$~cm$^{-3}$
and face moderate UV fields ($\chi \approx 10^3$, see Sect.~\ref{sect_model}).

\subsection{Mon~R2}

\subsubsection{\CII{}}

\begin{figure}
\resizebox{\hsize}{!}{\includegraphics[angle=90]{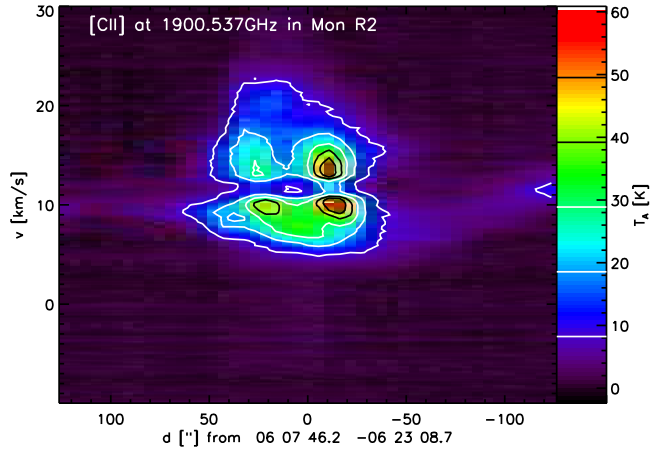}}
\caption{Position-velocity diagram of the \CII{} line for the measured cut through the
Mon~R2 PDR.}
\label{fig_monr2_ciipv}
\end{figure}

The \CII{} OTF map of Mon~R2 consists of a single strip across the
ultracompact \HII{} region from north-east to south-west \citep[see][]{fuente2010monr2}. In 
Fig.~\ref{fig_monr2_ciipv} we show the position-velocity diagram
for that cut. This is important for understanding the origin of the
components of the spectral profile in the comparison with the
\thirteenCII{} lines. We find a ring-like structure in position-velocity
space, with two well separated components towards the center of the
\HII{} region. A better understanding of the position velocity structure 
can be obtained by combining this information with equivalent maps
in several other tracers \citep{Pilleri2012}. In particular
the comparison with water lines, that show up in absorption in the
blue component but in emission in the red component, suggests that the
diagram traces a PDR around the \HII{} region that is accelerated by the 
radiative pressure from the central source, so that
the back side is red-shifted and the front blue shifted relative to the
systemic velocity of the medium. However, the observations show
the existence of some asymmetry between the expanding and receding layer
and part of the central dip
could also stem from self-absorption in the optically thick line.

\subsubsection{\thirteenCII{}}

The dual-beam-switch observations of the \thirteenCII{} lines were
performed at the center position of the OTF strip (position 0 in
Fig.~\ref{fig_monr2_ciipv}), i.e. not at the peak of the \CII{} 
emission. Future follow-up observations at the peak position will
provide a better accuracy.

\begin{figure}
\resizebox{\hsize}{!}{\includegraphics[angle=90]{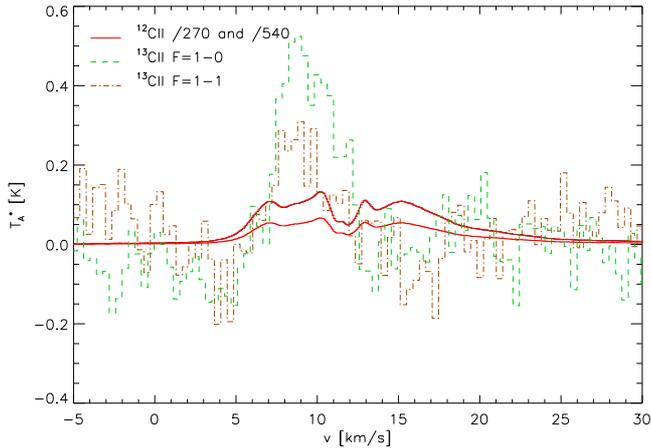}}
\caption{Comparison of the profiles of the \thirteenCII{}
hyperfine lines in Mon~R2 with the \CII{} profile in the same spectrum
scaled by the factors $0.25/67$ and $0.125/67$. {\changed The \thirteenCII{}
F=2-1 component is blended with the main isotopic line so that is not
shown here.}}
\label{fig_monr2_13cii}
\end{figure}

Figure~\ref{fig_monr2_13cii} shows the line profiles of the 
\thirteenCII{} components compared to the corresponding \CII{} spectrum,
scaled by the factor $0.25/67$ and $0.125/67$ indicating the expected
strength of the $F=1-0$ and $F=1-1$ hyperfine components in case of
optically thin emission and a canonical abundance ratio (see 
Sect.~\ref{sect_spectroscopy}).
Due to the large width of the \CII{} line, the $F=2-1$ {\changed component} falls in
the wing of the \CII{} emission, so that we cannot estimate its strength.
The striking result for the other two transitions is the
large difference in the spectral shape compared to the main isotopic 
line. The two hyperfine components are consistent with each other
in terms of the spectral shape, peaking at about 9~\kms{} and
in terms of the intensity ratio of about a factor two. The peak
agrees with a flat-top emission part of the \CII{} line that
would be characteristic for optically thick emission and is
about three times brighter than the scaled blue \CII{}
component. This is not reflected in the average optical depth 
$\langle \tau \rangle$ derived from the integrated line 
in Table~\ref{tab_intensities} due to the strong {\changedii mutual
difference} in the 
line shapes. To get an estimate for the two main velocity components, 
we have split the integration interval at $12.5$~\kms{} and
include the corresponding results in Table~\ref{tab_intensities}.
This increases the relative uncertainties, but allows to derive
independent quantitative results for the components that show
a very different behavior. For the blue-shifted foreground component,
we find the highest \CII{} optical depth in our sample. 
 
The big surprise is the complete absence of redshifted emission
from the back of the \HII{} region in \thirteenCII{}.  The {\changed red-shifted} \CII{} line
has approximately the same brightness as the blue component, but 
any possible \thirteenCII{} emission must be weaker than
expected from optically thin emission and normal abundance ratios.
For both hyperfine components an emission at the 0.1~K level can be 
excluded in spite of the baseline uncertainties.
{\changed 
This indicates the direct observation of fractionation
in a PDR driving an enhanced $^{12}$C$^+$/$^{13}$C$^+$ ratio. 
However, when considering the extreme uncertainties in 
Table~\ref{tab_intensities}, stemming from all possible baseline 
subtractions and the possibility that the redshifted \CII{} line
might be optically thin while the blueshifted line has an
optical depth $>3$, the error bars for the $^{12}$C$^+$/$^{13}$C$^+$
fractionation ratio still cover the standard elemental isotopic ratio so that even this case is not unambiguous.}

\subsection{NGC~3603, Carina, and NGC~7023}

\begin{figure}
\includegraphics[angle=90,width=8.3cm]{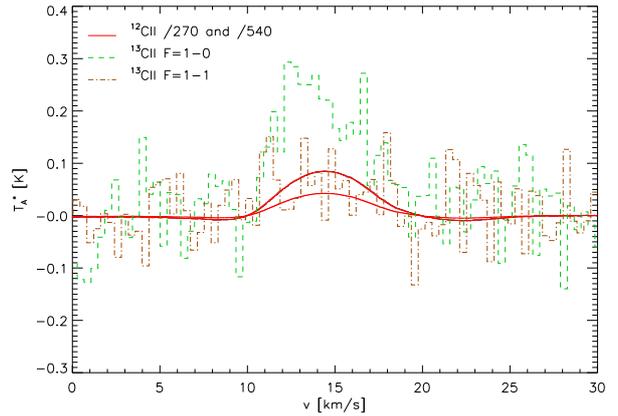}\\
\includegraphics[angle=90,width=8.3cm]{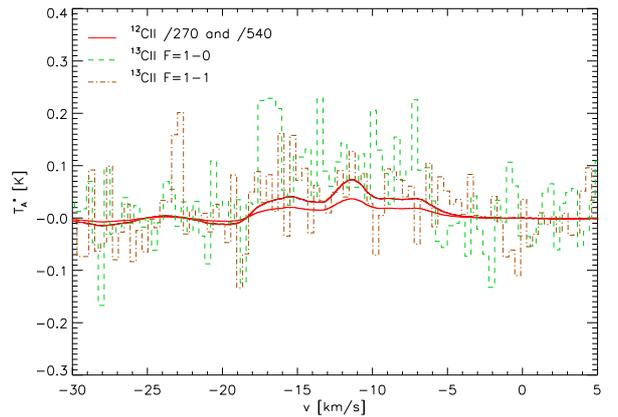}\\
\includegraphics[angle=90,width=8.3cm]{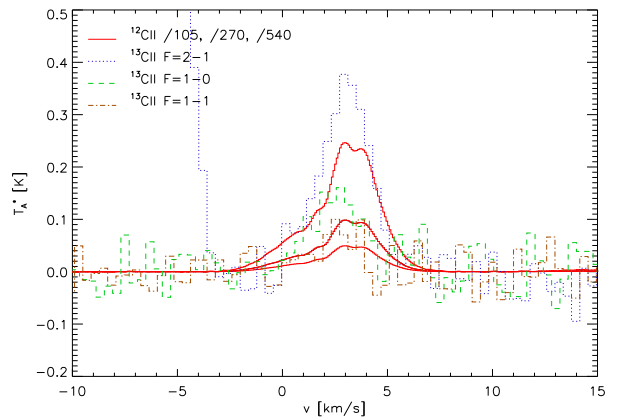}
\caption{Comparison of the profiles of the \thirteenCII{} hyperfine 
lines in NGC~3603~MM2 (upper panel), the Carina North PDR (central panel),
and NGC~7023 North (lower panel) with the \CII{} profiles from the same 
positions scaled by the factors corresponding to optically thin emission
and the standard abundance ratio for the resolved hyperfine components.
{\changed For NGC~3603~MM2 and Carina North, the \thirteenCII{}
F=2-1 component falls into a self-chopping feature of the main isotopic 
line so that is not shown here.}}
\label{fig_3603_carina}
\end{figure}

Figure~\ref{fig_3603_carina} shows the spectra for NGC~3603, 
Carina North, and NGC~7023, {\changed obtained in the same way as
Fig.\ref{fig_monr2_13cii} for Mon~R2}.
As discussed in Sect.\,\ref{sect_observations}, in NGC~3603 and Carina
the $F=2-1$ component falls into a \CII{} spectral feature resulting
from self-chopping in the observation, so
that it cannot be resolved. The two weaker components are resolved.
There is no indication for a spectral difference with respect to the 
\CII{} line and the mutual line ratio reflects the expected value from 
Sect.~\ref{sect_spectroscopy}. In both sources, the \thirteenCII{}
lines are about 2.5 times stronger than expected from optically thin \CII{} 
emission and the normal abundance ratio. The increase indicates an 
optical depth in the main isotopic line of $2.2-2.4$. 

The equivalent observation of the Carina South PDR, provided 
no detections of \thirteenCII{}. At the frequency of the $F=2-1$ transition
we find a weak emission component, but we rather attribute this to a separate
\CII{} velocity component in the complex velocity profile. As the \CII{} 
line brightness was more than three times weaker than in the other sources,
the non-detection is consistent with the expected line strength
in the optically thin case or cases with a moderate \thirteenCII{} line
enhancement as in the other sources.

The narrow \CII{} line in NGC~7023 is well separated from the \thirteenCII{}
$F=2-1$ component, so that this is the only observation {\changed in which} we resolve
all three hyperfine components. Moreover the narrow lines provide
a very small uncertainty for the line intensity from the baseline subtraction
so that we get quite accurate relative intensities.
 The average optical depth enhancement
of the \thirteenCII{} lines is moderate, but we also find a significantly
different behavior of different velocity components resolved in the
\CII{} line. The \thirteenCII{} lines mainly trace the component at 2.8~\kms{}
corresponding to the molecular material in the PDR \citep[see e.g.][]{Gerin1998}.
Here, we find a significant optical depth of the \CII{} line in terms of a
decreased \CII{}/\thirteenCII{} intensity ratio. The component at 4~\kms{} was identified as separate
filament by \citet{Fuente1996} and associated with ionized
material in the cavity by \citet{Joblin2010}. It is also seen in 
\thirteenCII{} {\changed and shows an} intensity ratio corresponding
to the elemental isotopic ratio.
Finally there is a high velocity wing with velocities around 0~\kms{}.
{\changedi This component may show an increased $^{12}$C$^+$/$^{13}$C$^+$
fractionation ratio through \thirteenCII{} intensities below the
scaled \CII{} intensity for all three hyperfine components.
However, our signal-to-noise ratio is not good enough to derive a 
definite statement on the \CII{}/\thirteenCII{} intensity ratio}
in that wing. Due to the almost Gaussian velocity components, the estimate 
of the C$^+$ column density from the integrated
line parameters should be a reasonable average by slightly underestimating the
optical depth from the PDR and slightly overestimating the contribution from the 
ionized gas in the cavity. For a total column density 
$N\sub{H} \approx 1.3\times 10^{22}$~cm$^{-2}$ at the position of our
observation \citep{Joblin2010} this indicates that more than half of
the carbon in the gas phase is in the form of C$^+$.

\section{Discussion}
\label{sec-discussion}

{\changed The detection of carbon fractionation under dense cloud
conditions remains difficult.} For the two species involved in
the most important carbon fractionation reaction, C$^+$ and CO, the
lines of the main isotopologue are often optically thick with a
considerable uncertainty in the optical depth. {\changed Consequently,
it is hard to determine the absolute column density and the
CO$/^{13}$CO or C$^+/^{13}$C$^+$ ratio. Our observations  and
simulations have confirmed this general knowledge for the 
C$^+/^{13}$C$^+$ ratio.} Observations of $^{13}$C$^{18}$O
are a possible way out, but challenging due to the low line
intensities, biased towards high column densities, and they
cannot trace the impact of isotope-selective photodissociation
provided by the self-shielding of the main isotopologue.

{\changed Our model predictions for PDRs} indicate that the observed fractionation ratio
in all tracers is dominated by the chemical fractionation reaction
(\ref{13eq1}) while isotopic-selective photo-dissociation plays a
minor role.  The {\changed chemical} fractionation {\changed increases} 
the C$^+/^{13}$C$^+$ ratio relative to the {\changedi elemental}
isotopic ratio in the gas. In terms of the observable
\CII{}/\thirteenCII{} ratio, this could lead to an increase by
a factor of a few under favorable conditions. They can be provided
by inner layers of moderate density, moderate UV PDRs
or by dense gas that is exposed to low UV fields
{\changed like clumps in an inhomogeneous PDR, embedded in 
and partially shielded by a thinner interclump medium.
The main condition for an observable increase of the
\CII{}/\thirteenCII{} ratio is, however, that}
geometry and velocity distribution prevent a large \CII{} 
optical depth. {\changed In most geometries,} the conditions providing an increased isotopic
ratio are paired with high optical depths of the \CII{} line
that rather reduces the \CII{}/\thirteenCII{} intensity ratio below the canonical value.

Our observations show that most of the \CII{} emission
must stem from {\changedii a widely distributed, low-density component in PDRs}, being
more smoothly distributed than many high-density tracers. Only
in Mon~R2, where the \HII{} region is still expanding into the
surrounding high-density core, a {\changed probably relatively homogeneous}
high-density
PDR is formed. The observational results on the \CII{}/\thirteenCII{} 
intensity ratio roughly follow the predictions from the models in Sect.\ref{sect_model}
without providing enough statistics for a qualitative confirmation yet.
For most of the PDRs, the average \CII{}/\thirteenCII{} intensity ratio
is {\changed reduced compared to the isotopic ratio}
due to the optical thickness of the \CII{} line. Details strongly depend
on the exact geometry. {\changed Assuming the standard elemental ratio for the
C$^+/^{13}$C$^+$ abundance ratio, we can measure the optical depth and consequently the C$^+$
column density through the intensity ratio.} The optical depth of the main isotopic line 
varies considerably across the line profile and between the different
PDRs, showing peak values above three for the Orion Bar and Mon~R2.

Indications for an increased \CII{}/\thirteenCII{} intensity ratio are found when studying details of the
line profiles allowing to distinguish different velocity components in
the same source. {\changed In three cases, we find components showing 
C$^+$ fractionation in a PDR. Unfortunately, none of them allows to
unambiguously quantify the degree of fractionation due to the large
error bars from the noise and baseline uncertainties (see
Table~\ref{tab_intensities}).
The clearest case} is the non-detection of \thirteenCII{} in the 
{\changed red-shifted} component in Mon~R2. If we interpret the red component
as the receding inner backside of the expanding \HII{} region, it represents a
dense, face-on PDR layer which should be at least partially optically thick in
\CII{}. As the front side shows the same integrated \CII{} intensity,
but no fractionation and a very high \CII{} optical depth instead, the
actual configuration must be asymmetric. {\changed The situation if even more
puzzling as our models predict negligible fractionation for this high-UV field
PDR.} Further high-resolution observations
are required to understand why the red-shifted material differs from the 
front material.
{\changed The second case is provided by the wings of the line profiles in the 
Orion Bar,} probably created by diffuse gas with a broader velocity distribution.
It may represent extended, possibly diffuse \HII{} gas in the Orion veil.
To confirm this origin, deeper integrations in positions apart from the
Orion Bar are needed.
A final case may be given by the blue wing material seen in NGC~7023.
However, in all three cases the step from a pure detection of fractionation
to a full quantitative assessment requires observations with a 
somewhat better signal-to-noise ratio than obtained here.

\begin{acknowledgements}
This work was supported by 
the \emph{Deut\-sche For\-schungs\-ge\-mein\-schaft, DFG\/} project
number Os~177/1--1 and SFB~956~C1. Additional support for this paper
was provided 
by NASA through an award issued by JPL/Caltech and by Spanish program 
CONSOLIDER INGENIO 2010, under grant CSD2009-00038 
Molecular Astrophysics: The Herschel and ALMA Era (ASTROMOL).
We thank Paul Goldsmith and an anonymous referee for many helpful
comments.

HIFI has been designed and built by a consortium of institutes and university
departments from across Europe, Canada and the US under the leadership of
SRON Netherlands Institute for Space Research, Groningen, The Netherlands
with major contributions from Germany, France and the US. Consortium
members are: Canada: CSA, U.Waterloo; France: CESR, LAB, LERMA,
IRAM; Germany: KOSMA, MPIfR, MPS; Ireland, NUI Maynooth; Italy: ASI,
IFSI-INAF, Arcetri-INAF; Netherlands: SRON, TUD; Poland: CAMK, CBK;
Spain: Observatorio Astron\'omico Nacional (IGN), Centro de Astrobiolog\'ia
(CSIC-INTA); Sweden: Chalmers University of Technology - MC2, RSS \&
GARD, Onsala Space Observatory, Swedish National Space Board, Stockholm
University - Stockholm Observatory; Switzerland: ETH Z\"urich, FHNW; USA:
Caltech, JPL, NHSC.
\end{acknowledgements}

\bibliographystyle{aa}
\bibliography{ref}
\end{document}